\documentclass{jfm}
\usepackage{graphics}
\usepackage{epstopdf, epsfig}

\usepackage{multiJFMequations}
\usepackage{amsmath,bm}
\usepackage{color}
\newcommand*{\be}{\begin{equation}}
\newcommand*{\ee}{\end{equation}}
\newcommand*{\bse}{\begin{subequations}}
\newcommand*{\ese}{\end{subequations}}
\newcommand*{\bme}{\begin{multiequations}}
\newcommand*{\eme}{\end{multiequations}}
\newcommand*{\se}{\singleequation}
\newcommand*{\de}{\doubleequation}
\newcommand*{\te}{\tripleequation}

\def\xcheck#1{\overset{\ifx#1f\hspace{.5ex}\fi\lower0.8ex\hbox{\tiny$\vee$}}{#1}}
\def\xbreve#1{\overset{\ifx#1f\hspace{.5ex}\fi\lower0.8ex\hbox{\tiny$\smile$}}{#1}}
\def\xmathring#1{\overset{\ifx#1f\hspace{.5ex}\fi\lower0.8ex\hbox{\tiny$\circ$}}{#1}}

\newcommand*{\mapsfrom}{\mbox{\reflectbox{$\mapsto$}}}
\newcommand*{\sumpm}{\hbox{$\sum_\pm$}}

\newcommand*{\XXint}[3]{{\setbox0=\hbox{$#1{#2#3}{\int}$}
\vcenter{\hbox{$#2#3$}}\kern-.5\wd0}}

\newcommand*{\XXnotinfty}[3]{{\setbox0=\hbox{$#1{#2#3}{\to}$}
\vcenter{\hbox{$#2#3$}}\kern-.5\wd0}}

\newcommand*{\ds}{\displaystyle}
\providecommand*{\dfrac}[2]{\ds\frac{#1}{#2}}

\renewcommand*{\Im}{\mbox{Im}}

\renewcommand*{\tilde}{\widetilde}
\renewcommand*{\hat}{\widehat}
\renewcommand*{\bar}{\overline}

\newcommand*{\od}[2]{\dfrac{{\mathrm d}{#1}}{{\mathrm d}{#2}}}
\newcommand*{\pd}[2]{\dfrac{\partial{#1}}{\partial{#2}}}

\newcommand*{\erf}{\,{\mbox{erf}}\,}
\newcommand*{\erfc}{\,{\mbox{erfc}}\,}

\renewcommand*{\Lambda}{\varLambda}
\renewcommand*{\Upsilon}{\varUpsilon}
\renewcommand*{\Phi}{\varPhi}
\renewcommand*{\Psi}{\varPsi}
\renewcommand*{\Omega}{\varOmega}
\renewcommand*{\Theta}{\varTheta}
\renewcommand*{\Xi}{\varXi}
\newcommand*{\Omegav}{{\bm{\Omega}}}
\newcommand*{\dR}{{\mathrm d}}\newcommand*{\iR}{{\mathrm i}}
\newcommand*{\CR}{{\mathrm C}}\newcommand*{\IR}{{\mathrm I}}\newcommand*{\JR}{{\mathrm J}}\newcommand*{\SR}{{\mathrm S}}
\newcommand*{\vv}{{\bm{v}}}
\newcommand*{\dS}{{\sf{d}}}
\newcommand*{\CS}{{\sf{C}}}\newcommand*{\ES}{{\sf{E}}}\renewcommand*{\SS}{{\sf{S}}}
\newcommand*{\CC}{{\mathcal C}}\newcommand*{\EC}{{\mathcal E}}\newcommand*{\LC}{{\mathcal L}}\newcommand*{\SC}{{\mathcal S}}
\newcommand*{\pG}{{\mathfrak p}}\newcommand*{\sG}{{\mathfrak s}}
\newcommand*{\CG}{{\mathfrak C}}\newcommand*{\EG}{{\mathfrak E}}\newcommand*{\FG}{{\mathfrak F}}\newcommand*{\HG}{{\mathfrak H}}\newcommand*{\RG}{{\mathfrak R}}\newcommand*{\WG}{{\mathfrak W}}\newcommand*{\ZG}{{\mathfrak Z}}
\newcommand*{\vt}{{\tilde v}}

\newcommand*{\chit}{{\tilde \chi}}

\newcommand*{\chimr}{{\mathring \chi}}\newcommand*{\vmr}{{\mathring v}}

\newcommand*{\ub}{{\bar u}}\newcommand*{\vb}{{\bar v}}

\newcommand*{\chib}{{\bar\chi}}

\newcommand*{\vth}{{\hat \vt}}\newcommand*{\chith}{{\hat \chit}}
\newcommand*{\vmrh}{{\hat \vmr}}\newcommand*{\chimrh}{{\hat \chimr}}

\newcommand*{\vvh}{{\hat \vv}}

\newcommand*{\ubh}{{\hat \ub}}

\newcommand*{\EGh}{{\hat \EG}}\newcommand*{\WGh}{{\hat \WG}}

\newcommand*{\chixw}{{\xcheck \chi}}
\newcommand*{\vxw}{{\xcheck v}}
\newcommand*{\chixbr}{{\xbreve \chi}}
\newcommand*{\vxbr}{{\xbreve v}}

\newcommand*{\tE}{{\mbox{\tiny {\itshape E}}}}

\newcommand*{\tIW}{{\mbox{\tiny {\itshape IW}}}}
\newcommand*{\tGH}{{\mbox{\tiny {\itshape MF}}}}
\newcommand*{\tQG}{{\mbox{\tiny {\itshape QG}}}}

\newcommand*{\tDNS}{{\mbox{\tiny {DNS}}}}
\newcommand*{\tFNS}{{\mbox{\tiny {FNS}}}}

\newcommand*{\tEG}{{\mbox{\tiny {$\EG$}}}}
\newcommand*{\tWG}{{\mbox{\tiny {$\WG$}}}}

\renewcommand{\star}{{\dag}}

\newcommand*{\OSD}[1]{(I:\,{#1})}

\newcommand*{\Ro}{{Ro\,}}

\shorttitle{Spin-down inertial waves. Part II}
\shortauthor{L. Oruba, A.~M. Soward and E. Dormy}

\title{The inertial wave activity during spin-down in a rapidly penny shaped cylinder.\\
Part II The inertial wave of maximum frequency trigger}

\author{L. Oruba \aff{1}$^{,}\dag$
  A. M. Soward \aff{2}$^{,}$\corresp{\email{ludivine.oruba@latmos.ipsl.fr, andrew.soward@ncl.ac.uk, Emmanuel.Dormy@ens.fr}},
 \and E. Dormy\aff{3}$^{,}\dag$}

\affiliation{\aff{1} Laboratoire Atmosph\`eres Milieux Observations Spatiales (LATMOS/IPSL), Sorbonne Universit\'e, UVSQ, CNRS, Paris, FRANCE
\aff{2} School of Mathematics and Statistics, Newcastle University, Newcastle upon Tyne NE1 7RU, UK
\aff{3} {D\'epartement de Math\'ematiques et Applications, UMR-8553, \'Ecole Normale Sup\'erieure, CNRS, PSL University, 75005 Paris, FRANCE}}

\begin{document}

\maketitle
%    \centerline{\today}

\begin{abstract}
  In an earlier paper, Oruba, Soward \& Dormy (J.~Fluid Mech., vol.~818, 2017, pp.~205--240) considered the primary quasi-steady geostrophic (QG) motion of a constant density fluid of viscosity $\nu$ that occurs during linear spin-down in a cylindrical container of radius $L$ and height $H$, rotating rapidly (angular velocity $\Omega$) about its axis of symmetry subject to mixed rigid and stress-free boundary conditions for the case $L=H$. Direct Numerical Simulation (DNS) at large $L= 10 H$ and Ekman number $E=\nu/H^2\Omega=10^{-3}$ by Oruba, Soward \& Dormy (J.~Fluid Mech., sub judice and referred to as Part I) reveals significant inertial wave activity on the spin-down time-scale. The analytic study in Part~I, based on $E\ll 1$, builds on the results of Greenspan \& Howard (J.~Fluid Mech., vol.~17, 1963, pp.~385--404) for an infinite plane layer $L\to\infty$. At large but finite distance $r^\dag$ from the symmetry axis, the meridional (QG-)flow, that causes the QG-spin down, is blocked by the lateral boundary $r^\dag=L$, which provides the primary QG-trigger for the inertial waves studied in Part~I. For the laterally unbounded layer, Greenspan \& Howard also identified inertial waves of maximum frequency (MF), which are  a manifestation of the transient Ekman layer. The blocking of the MF-waves by the lateral boundary provides a secondary MF-trigger for yet more inertial waves. Here we obtain analytic results for the wave activity caused by the combined-trigger (QG+MF) that faithfully captures the character of the laterally unbounded base flow including its transients. The results are compared with the inertial wave part of the DNS (the so called ``filtered DNS'' or simply ``FNS''), for which the agreement is excellent and accounts for minor discrepancies evident in the Part~I results for the  QG-trigger.
\end{abstract}

%%%%%%%%%%%%%%%%%%%%%%%%%%%%%%%%%%%%%
%%%%%%%%%%%%%%%%%%%%%%%%%%%%%%%%%%%%%
%%%%%       SECTION 1
%%%%%%%%%%%%%%%%%%%%%%%%%%%%%%%%%%%%%
%%%%%%%%%%%%%%%%%%%%%%%%%%%%%%%%%%%%%

\section{Introduction\label{Introduction}}

In this paper, we continue our  Part~I investigation in \citet{OSD18} of the inertial wave response during spin-down in a shallow cylinder height $H$, radius $L=\ell H$,
\be
\label{L-large}
L\gg H  \hskip 10mm \mbox{equivalently}\hskip 10mm \  \ell \gg 1
\ee
As we need to refer extensively to equations (say ($x$.$y$)), sections (say {\S}$x$) and figures (say figure~$x$) from Part~I, we use the notation ``\OSD{$x$.$y$}'', ``\S I:$x$'' and ``figure~I:$x$'' respectively to identify them.

Our cylindrical container is filled with constant density fluid of viscosity $\nu$ and rotates rigidly with angular velocity $\Omegav$ about its axis of symmetry, the frame, relative to which our analysis is undertaken; the Ekman number is small:
\be
\label{Ek-numb}
E\,=\,\nu \big/\bigl(H^2\Omega\bigr)\,\ll\, 1\,.
\ee
Initially, at time $t^\star=0$, the fluid itself rotates rigidly at the slightly larger angular velocity $\Ro\Omegav$, in which the Rossby number $\Ro\!$ is sufficiently small ($\Ro\!\ll E^{1/4}$) for linear theory to apply. Relative to cylindrical polar coordinates, $(r^\star,\,\theta^\star,\,z^\star)$, the top boundary ($r^\star<L$, $z^\star=H$) and the side-wall ($r^\star=L$, $0<z^\star<H$) are impermeable and stress-free. The lower boundary ($r^\star<L$, $z^\star=0$) is rigid. For that reason alone the initial state of relative rigid rotation  $\Ro\Omegav$ of the fluid cannot persist and the fluid spins down to the final state of no rotation relative to the container, as $t^\star\to\infty$. In order to make our notation relatively compact at an early stage, we use $H$ and $\Omega^{-1}$ as our unit of length and time respectively, and introduce 
\bme
\label{dim-length-time}
\be
\te
r^\star=Hr\,,\qquad z^\star=Hz\,,\qquad\qquad  \Omega t^\star=t\,.
\ee
\eme
For our unit of relative velocity $\vv^\star$, we adopt the velocity increment $\Ro L \Omega$ of the initial flow at the outer boundary $r^\star=L$. So, relative to cylindrical components, we set
\bme
\label{vel}
\be
\vv^\star\,=\,\Ro L \Omega\,\vv\,, \qquad\qquad  \vv\,=\,[u,\,v,\,w]
\ee
and introduce the streamfunction $r\chi$ for the meridional flow:
\be
u\,=\,-\,\pd{\chi}{z}\,,\hskip 15mm w\,=\,\dfrac{1}{r}\pd{(r\chi)}{r}\,.
\ee
\eme
Throughout this paper, our investigation of the transient solution will rely heavily on the Laplace transform (LT: an operation $\LC$ that we denote by the $\,\,\hat{}\,\,\,$ accent), e.g.,
\bse
\label{vel-LT}
\begin{align}
\vvh(r,z,p)\,=\,\LC_p\{\vv\}\,\equiv\,&\,\int_0^\infty \vv(r,z,t)\,\exp(-pt)\,\dR t\,,
\intertext{where the subscript `$p$' to $\LC$ identifies the independent transform variable. The inverse-LT is  \vskip -3mm}
 \vv(r,z,t) \,=\,\LC^{-1}_p\{\vvh\} \,\equiv\,&\,\dfrac{1}{2\pi \iR}\int_{-\iR\infty}^{\iR\infty}\vvh(r,z,p)\exp(pt)\,\dR p\,.
\end{align}
\ese

\subsection{The QG, MF~and combined-triggers\label{QG-MF-combined-triggers}}

Following Part~I, we build on the study of \cite{GH63} for an unbounded plane layer $\ell \to \infty$. The essential idea is that its solution provides a first approximation to the bounded case of $\ell$ large but finite. The usefulness of the unbounded layer solution lies in the fact that, for a long period of time
\bse
\label{Et-small}
\be
Et\,\ll\,1\,,
\ee
the $[u,v]$-motion is $z$-independent outside boundary layers. Importantly, an expanding diffusion layer of width
\be
\Delta(t)\,=\,\sqrt{Et}
\ee
\ese
\OSD{1.17$b$} forms on the lower boundary $z=0$. The quasi-steady Ekman layer, width $\Delta_\tE\equiv\Delta(1)=E^{1/2}$, is established on the rotation time $t=O(1)$, after which its  primary role is to spin-down the quasi-geostrophic (QG-)flow ${\bar \vv}_{\tQG}(r,t)$ (say) above it, $\Delta_\tE\ll z\le 1$. Subsequently, for $t\gg 1$, the transient (decaying) shear layer continues to thicken, $z=O(\Delta(t))$, and drives $z$-independent inertial waves of maximum frequency (MF) $2$, velocity ${\bar \vv}_{\tGH}(r,t)$:
\bme
\label{chib-MF}
\be
\left[\begin{array}{c}  \!\! \ub_\tGH \!\!\\[0.2em]
    \!\!  \vb_\tGH \!\!  \end{array}\right]
\approx\,\dfrac{r}{\ell}\,\dfrac{E^{1/2}}{\sqrt{4\pi t}}\left[\begin{array}{c}  \!\! -\cos(2t) \!\!\\[0.2em]
    \!\!  \sin(2t)\!\!  \end{array}\right], \hskip 12mm
\chib_\tGH\,\approx\,\dfrac{r}{\ell}\,\dfrac{E^{1/2}}{\sqrt{4\pi t}}\,(z-1)\cos(2t)\,,
\ee
\eme
in the mainstream outside, $(\Delta_\tE\ll) \,\Delta(t)\ll z \le 1$. The complete MF-flow $\vv_{\tGH}(r,z,t)$ is composed of mainstream (${\bar \vv}_{\tGH}$) and shear layer parts, which together are
\bme
\label{MF-combined}
\se
\begin{align}
\left[\begin{array}{c}  \!\! u_\tGH \!\!\\[0.2em]
    \!\!  v_\tGH \!\!  \end{array}\right]
\approx\,&\,\dfrac{r}{\ell}\,\dfrac{E^{1/2}}{\sqrt{4\pi t}}\left[\begin{array}{c}  \!\! -\cos(2t) \!\!\\[0.2em]
    \!\!  \sin(2t)\!\!  \end{array}\right]\biggl[1-\dfrac{z}{2Et}\,\exp\biggl(-\dfrac{z^2}{4Et}\biggr)\biggr],\\[0.3em]
\chi_\tGH\,\approx\,&\,\dfrac{r}{\ell}\,\dfrac{E^{1/2}}{\sqrt{4\pi t}}\,\cos(2t) \biggl[z-1+\exp\biggl(-\dfrac{z^2}{4Et}\biggr)\biggr]
\end{align}
\eme
\OSD{1.27}. Clearly this mainstream and boundary layer partition relies on $\Delta(t)\ll 1$, a condition that is met when $Et\ll 1$ (whence the restriction (\ref{Et-small}$a$)).

In Part~I, we investigated the response for $r<\ell$ caused by blocking  the primary radial QG-velocity
\bme
\label{uQG-L}
\be
\ub_{\tQG}(\ell,t)\,\approx\, \tfrac12 \sigma\kappa E^{1/2}\,\EG(t)\,, \hskip15mm    \EG(t)\,=\,\exp(-E^{1/2}\sigma t)\,,
\ee
of the unbounded flow at $r=\ell$, $0<z\le 1$ ($\sigma\approx  1+\tfrac34 E^{1/2}$, $\kappa\sigma\approx  1+E^{1/2}$: see \OSD{1.18$c$-$e$}, \OSD{1.30$a$}). The LT of $\EG(t)$ is
\be\se
\EGh(p)\,=\,\bigl(p+E^{1/2}\sigma\bigr)^{-1}\,.
\ee
\eme
Here, by contrast, we wish to consider the additional response due to radial component of the mainstream part of the secondary MF-flow (\ref{chib-MF}) at $r=\ell$, namely
\bse
\label{uMF-L}
\be
\ub_{\tGH}(\ell,t)\,\approx\, -\,\dfrac{E^{1/2}}{\sqrt{4\pi t}}\,\cos(2 t) \hskip 10mm \mbox{for} \hskip 7mm t\gg 1
\ee
with LT
\be
\ubh_{\tGH}(\ell,p)\,\approx\,-\,\tfrac14 E^{1/2}\,\Bigl[\bigl(p+2\iR\bigr)^{-1/2}\,+\,\bigl(p-2\iR\bigr)^{-1/2}\Bigr].
\ee
\ese

However,  \cite{GH63} suggested on their pp. 390, 391, that under the approximation $\sigma\kappa\approx 1$, a uniformly valid approximation for $0<t\ll E^{-1}$ of the combined motions $\ub_{\tQG}(r,t)$ and $\ub_{\tGH}(r,t)$ is provided by $(\ell/r)\ub_\tWG(r,t)=-\partial {\phi_I}/\partial z$, where $\phi_I(r,t)$ is defined by their eq.~(3.17). In our notation it is
\bme
\label{u-entire}
\be
\ub_\tWG(r,t)\,=\,\tfrac12 E^{1/2}(r/\ell)\,\WG(t) \hskip 15mm \mbox{with} \hskip 15mm \WGh(p)\,=\,\EGh(p)\!\;\RG(p)\,,
\ee
where
\be\se
\RG(p)\,=\,\iR\Bigl[\bigl(p+2\iR\bigr)^{-1/2}\,-\,\bigl(p-2\iR\bigr)^{-1/2}\Bigr].
\ee
At $r=\ell$, $\ub_\tWG$ takes the value
\be\se
\ub_\tWG(\ell,t)\,\approx\,\tfrac12 E^{1/2}\,\WG(t)\,.
\ee
\eme
From this viewpoint, it is convenient to replace the definition (\ref{uMF-L}$a$) of $\ub_{\tGH}(\ell,t)$ by
\bme
\label{MF-towards-unified}
\be
\ub_{\tGH}(\ell,t)\,=\,\tfrac12 E^{1/2}\WG_{\tGH}(t)\,, \hskip 20mm \WG_{\tGH}(t)\,=\,\WG(t)-\EG(t)\,,
\ee
where, instead of the LT (\ref{uMF-L}$b$), we now have
\be\se
\WGh_{\tGH}(p)\,=\,\EGh(p)\bigl[\RG(p)-1\bigr].
\ee
\eme
For $t\gg 1$, the asymptotic evaluation of the inverse-LT of (\ref{MF-towards-unified}$c$) is dominated by the cut contributions near $p=\pm2 \iR$, which coincide with those of (\ref{uMF-L}$b$). By implication, correct to leading order, the asymptotic form of (\ref{MF-towards-unified}$a$) for $t\gg 1$ recovers (\ref{uMF-L}$a$). Be that as it may, the new definition has the advantage that
\be
\label{W=QG+MF}
\ub_\tWG(\ell,t)\,=\,\ub_{\tQG}(\ell,t)\,+\,\ub_{\tGH}(\ell,t)\,,\hskip 10mm \mbox{when} \hskip 10mm\sigma\kappa\approx 1\,,
\ee
for all $t>0$. We refer to the boundary conditions, $u(\ell,t)=-\ub_{\tQG}(\ell,t)$, $-\ub_{\tGH}(\ell,t)$ and $-\ub_{\tWG}(\ell,t)$, as the QG (or $\EG$-), MF- and combined (or $\WG$-)triggers respectively. Finally, we note that at $r=\ell$ the triggers are valid outside the respective boundary layers of the flows that define them, i.e., the QG-trigger on $\Delta_\tE\ll z\le 1$, the MF-trigger on $\mathrm{max}\{\Delta_\tE,\Delta(t)\}\ll z\le 1$, though for $t=O(1)$ the $\WG$-trigger fairs better than both on $\Delta(t)\ll z\le 1$. A comprehensive discussion of the nature of the MF-flow $\vv_\tGH(r,z,t)$ (included in (\ref{c-of-v}$b$) below) was given in \S{I:1.2.2}.

\subsection{An appraisal of the combined $\WG$-trigger and commentary on approximations\label{appraisal}}

The numerical results displayed by the figures in Part~I, were obtained on neglecting the exponential decay $\exp(-E^{1/2}\sigma t)$ (see \OSD{3.8}) of the QG-trigger (\ref{uQG-L}$b$), as it was found to have virtually no influence on the solution. Implementation involved approximating the pole of $\EGh(p)$ at $p=-E^{1/2}\sigma$ in (\ref{uQG-L}$c$), which identifies the spin-down time $E^{-1/2}\sigma$, by a pole at $p=0$. Here we adopt the same approximation 
\bme
\label{basic-approx}
\be
\EG(t)\,=\,1\,,\hskip 20mm \EGh(p)\,=\,p^{-1}\,,
\ee
\eme
which henceforth supersedes the definitions (\ref{uQG-L}$b$,$c$). Accordingly, the combined-trigger factor $\WG(t)$ (\ref{u-entire}$a$) has LT $\WGh(p)$ (\ref{u-entire}$b$), which may be expressed compactly as
\bse
\label{our-trigger-a}
\be
\WGh(p)\,=\,\sumpm\Bigl\{\WGh^\pm(p)\Bigr\}\,\equiv\,\WGh^+(p)\,+\,\WGh^-(p)\,,
\ee
where our notation $\sumpm\{\bullet\}$ means the two term, $+$ and $-$, sum, as illustrated, and
\be
\WGh^\pm(p)\,=\,\pm\iR p^{-1}\bigl(p\pm 2\iR\bigr)^{-1/2}\,.
\ee
\ese
Interestingly, this construction of $\WGh(p)$ coincides with \OSD{1.11$b$} for the transient Ekman layer in an otherwise unbounded fluid. Moreover, noting that $\ZG(t)$ of \OSD{1.9} is related to $\WG^-(t)$ by $-\,\ZG(t)= \WG^{-}(t)$, we may restate \OSD{1.9}--\OSD{1.11} in the form
\bse
\label{our-trigger-b}
\begin{align}
 \WG^{\pm}(t)\,=\,\bigl((\pm\iR)^{1/2}\big/\sqrt{2}\,\bigr)\,\erf\!\Bigl((\pm\iR)^{1/2}\sqrt{2t}\,\Bigr)=\,&\,\SR\bigl(2t^{1/2}/\pi^{1/2}\bigr)\pm\iR\CR\bigl(2t^{1/2}/\pi^{1/2}\bigr)\,,\\
\WG(t)\,=\,\sumpm\bigl\{\WG^\pm(t)\bigr\}\,=\,&\,2\!\;\SR\bigl(2t^{1/2}/\pi^{1/2}\bigr)\,,
\end{align}
where $\SR$ and $\CR$ are Fresnel integrals:
\be
\biggl[\begin{array}{c}  \!\! \SR(2t^{1/2}/\pi^{1/2}) \!\!\\[0.1em]
    \!\!\CR(2t^{1/2}/\pi^{1/2})   \!\!  \end{array}\biggr]\,
=\,\int_0^{t}
\biggl[\begin{array}{c}  \!\!  \sin 2\tau \!\!\\[0.1em]
    \!\!  \cos 2\tau \!\!  \end{array}\biggr]\,\,\dfrac{\dR \tau}{\sqrt {\pi\tau}}\,.
\ee
\ese
The initial behaviour ($t\ll 1$),
\bme
\label{soft-trigger}
\be
\WG(t)\,\approx\,\dfrac{8}{3\sqrt\pi}\:t^{3/2} \hskip 12mm \Longleftrightarrow\hskip 12mm \WGh(p)\,=\,\dfrac{2}{p^{5/2}}\,,
\ee
\eme
may be derived by the LT-inversion of the sum (\ref{our-trigger-a}$a$) subject to the restriction $|p|\gg 1$.

A nagging concern about our strategy is that, whereas the QG-trigger (\ref{uQG-L}$a$) involves the factor $\kappa\sigma\approx 1+E^{1/2}$, the uniformly valid combined $\WG$-trigger (\ref{u-entire}) is by necessity based on $\kappa\sigma\approx 1$. To assess any possible weakness in our Greenspan \& Howard starting point (\ref{u-entire}), we obtained results (not illustrated here) based on (\ref{uQG-L}$a$) for the QG-trigger together with the asymptotic  ($t\gg 1$) form (\ref{uMF-L}) for the MF-trigger. Some minor discrepancies with the $\WG$-trigger results were apparent, but, relative to the consequences of other approximations made, they were so small as to be of no concern.

To understand why some approximations (see particularly items (i), (ii) below), which superficially look suspect, seem to work well, we need to appreciate how the applied boundary condition at $r=\ell$ drives motion. As explained in Part~I, the dominant feature of the solution is a wave packet that travels inwards from the boundary $r=\ell$ at the group velocity (the wave itself propagates in the opposite direction outwards). What remains near $r=\ell$ exists on ever temporally decreasing length scales and so is subject to considerable dissipation by internal friction. From that perspective the early time behaviour near $r=\ell$ is largely responsible for the visible behaviour elsewhere at later times, an aspect that we emphasise by our description of the boundary condition as a ``trigger''.

The trigger property may explain why
\begin{itemize}
\item[(i)] the neglect of the slow decay $\exp(-E^{1/2}\sigma t)$ on the long spin-down scale in our construction (\ref{our-trigger-b}$b$) of the $\WG$-trigger seems to be of little consequence;
\item[(ii)] the application of the MF-trigger on the range $0< z\le 1$, rather than on the ever temporally shrinking domain $\Delta(t)\ll z\le 1$ on which the entire mainstream $\WG$-flow $\ub_\tWG(r,t)$ (\ref{u-entire}$a$) resides, appears to lead to good results, except in a relatively small region near the corner $(r,z)=(\ell,0)$. The significant point here is that the boundary layer width $\Delta(t)$ increases from $0$  at $t=0$ to $\Delta_\tE$ at $t=1$, and so on that early rotation time scale the application of the $\WG$-trigger on $0< z\le 1$ is indeed a very good approximation.
\end{itemize}

\subsection{Outline\label{Outline}}

The paper is organised as follows. In \S\ref{mathematical-problem}, we formulate and partially solve the mathematical problem for the inertial waves generated by the combined $\WG$-trigger. In \S\ref{no-dis}, we extract the solution for $E=0$, which must be understood in terms of the viscous solution in the limit $E\downarrow 0$ over the time-interval $0<t \ll E^{-1}$ (including the spin-down time $t=O(E^{-1/2})$) restricted to the mainstream region exterior to all boundary layers. In \S\ref{dis}, we consider the role of viscosity ($0< E \ll 1)$ in damping the inertial wave structures predicted in \S\ref{no-dis}. In Part~I, the $\EG$-trigger predictions, based on $\ub_{\tQG}(\ell,t)=\tfrac12 \sigma\kappa E^{1/2}$ (see \OSD{1.30$a$} with $\EG(t)=1$) were compared with the results derived from the Direct Numerical Simulations (DNS) of the governing equations subject to the complete set of initial and boundary conditions for the case $E=10^{-3}$, $\ell=10$. That study motivates the similar comparison in \S\ref{numerics} of our new $\WG$-trigger findings, based on $\ub_\tWG(\ell,t)=\tfrac12 E^{1/2}\,\WG(t)$ adopted in (\ref{G-bc}) below. Despite the difficulties associated with making sensible approximations in \S\ref{dis} to accommodate viscous damping, our $\WG$-trigger results of \S\ref{numerics} significantly improve agreement with the DNS. Though this tangible success was our key motivation, the  $E\downarrow 0$ results of \S\ref{no-dis} are significant because they are analytically robust and clearly identify the fine inertial wave structure generated by the $\WG$-trigger, which is heavily damped when  $E=10^{-3}$. We conclude with a brief overview in \S\ref{Discussion}.

%%%%%%%%%%%%%%%%%%%%%%%%%%%%%%%%%%%%%
%%%%%%%%%%%%%%%%%%%%%%%%%%%%%%%%%%%%%
%%%%%       SECTION 2
%%%%%%%%%%%%%%%%%%%%%%%%%%%%%%%%%%%%%
%%%%%%%%%%%%%%%%%%%%%%%%%%%%%%%%%%%%%

\section{The mathematical problem\label{mathematical-problem}}

Our strategy parallels Part~I and so here we only sketch the methodology;  for a more careful appraisal, the reader is referred to that work. The essential idea is that the flow $\vv_{\tWG}$ between unbounded parallel planes ($\ell\to \infty$), whose LT-solution is given by eqs.~(3.4)--(3.6) of \cite{GH63}, provides the lowest order solution to the bounded ($\ell$ large but finite) problem. The main point, emphasised in \S\ref{Introduction}, is that outside boundary layers the horizontal components $[\ub_{\tQG},\vb_{\tQG}]$ and $[\ub_{\tGH},\vb_{\tGH}]$ of both the QG and MF-flow contributions (for $\ell\to \infty$) are $z$-independent. However, the failure of the radial velocities $\ub_{\tQG}$ and $\ub_{\tGH}$, both of $O(E^{1/2})$ (see (\ref{uQG-L})-(\ref{u-entire})), to meet the requirement $\ub(\ell,t)=0$ triggers a further inertial wave response. As we are only interested in that response outside the Ekman and side-wall boundary layers, we write
\bme
\label{c-of-v}
\be
\vv\,\approx\,\vv_{\tWG}\,+\,E^{1/2}\vv^{\rm {wave}}\,, \hskip 10mm \vv_{\tWG}\,=\,{\bar \vv}_{\tQG}\,+\,\vv_{\tGH}\,,
\ee
\eme
in which $\vv_{\tGH}$ takes the asymptotic form (\ref{MF-combined}) for $t\gg 1$. Our objective is to determine $\vv^{\rm {wave}}$ obtained subject to the $\WG$-trigger boundary condition
\be
\label{G-bc}
u^{\rm {wave}}\,=\,-\,E^{-1/2}\ub_{\tWG}(\ell,t)\,=\,-\,\tfrac12\WG(t) \hskip 15mm \mbox{at} \hskip 6mm r=\ell   \hskip 6mm  (0<z\le 1) 
\ee
(see (\ref{u-entire}$d$) with (\ref{our-trigger-b}$b$) and cf.~\OSD{2.3} for the $\EG$-trigger).

Throughout this  section we drop the superscript `wave' and write $\vv=[u,\,v,\,w]$ ($\mapsfrom\,\vv^{\rm {wave}}$). With $w=r^{-1}\partial(r\chi)/\partial r$ (\ref{vel}$d$), the inertial wave problem is: Solve
\bme
\label{gov-eqs}
\begin{align}
\pd{v}{t}\,+\,2\,u\,&=\,E\bigl(\nabla^2-r^{-2}\bigr)v\,,&    u\,&=\,-\,\pd{\chi}{z}\,,\\[0.3em]
\pd{\gamma}{t}\, -\,2\,\pd{v}z\,&=\,E\bigl(\nabla^2-r^{-2}\bigr)\gamma\,,&
\gamma\,&=\,-\, \bigl(\nabla^2-r^{-2}\bigr)\chi
\end{align}
\eme
subject to the initial ($t=0$) conditions 
\bme
\label{initial-condits}
\be
v = 0,\qquad\qquad\qquad     \gamma=0,
\ee
\eme
and for $t\ge 0$ the boundary conditions 
\bse
\label{boundary-condits}
\begin{align}
  r\chi\,&=\,0& \mbox{at} &&r\,=&\,0\, & (0&<z\le 1)\,,\\
  r\chi\,&=\,\tfrac12\ell (z-1)\;\!\WG(t)& \mbox{at} &&r\,=&\,\ell\, & (0&<z\le 1)\,,\\ 
\chi\,&=\,0&  \mbox{at}&& z\,=&\,0,\,1 & (0&<r<\ell)\,.
\end{align}
\ese
This is the Part~I problem \OSD{2.4}--\OSD{2.6} but modified by the replacement $\kappa\sigma\EG\mapsto\WG$ in \OSD{2.6$b$}.

\subsection{The $z$-Fourier series \label{Fourier-series}}

We seek $z$-Fourier series solutions of the form 
\be
\label{GH-FS}
\biggl[\begin{array}{c}  \!\! \chi \!\!\\[0.3em]
    \!\!  v \!\!  \end{array}\biggr]\,=\,-\sum_{m=1}^\infty\dfrac{(-1)^m}{m \pi}
\biggl[\begin{array}{c}  \!\! \chit_m(r,t)\sin\bigl(m\pi (z-1)\bigr) \!\!\\[0.3em]
    \!\! \vt_m(r,t)\cos\bigl(m\pi (z-1)\bigr)  \!\!  \end{array}\biggr]
\ee
(see \OSD{2.8$a$,$b$} with $\kappa\sigma=1$) for which (\ref{boundary-condits}$b$), noting
\be
\label{z-minus-one}
\tfrac12 (z-1)\, = \,-\,\sum_{m=1}^\infty\dfrac{(-1)^m}{m\pi}\sin(m\pi (z-1)) \hskip 15mm    (0<z\le 1)
\ee
\OSD{2.7}, leads to the boundary condition
\be
\label{pulse-m-bc}
r\chit_m\,=\, \ell\;\!\WG(t) \hskip 12mm \mbox{at} \hskip 12mm r\,=\,\ell\,.
\ee
The LT-solution \OSD{2.16$a$} following the change $\EGh(p)\mapsto\WGh(p)$ is
\bme
\label{pulse-m-LT}
\be\se
\left[\begin{array}{c}  \!\! \chith_m \!\!\\[0.3em]
    \!\!  \vth_m \!\!  \end{array}\right]\,=
\left[\begin{array}{c}  \!\! 1 \!\!\\[0.3em]
      \!\!  2m\pi/\pG \!\!  \end{array}\right]\,\WGh(p)\,
\dfrac{\JR_1\bigl(m\pi qr\bigr)}{\JR_1\big(m\pi q\ell\bigr)}\,.
\ee
The dispersion relation \OSD{2.17$a$--$d$} gives
\be
\left.\begin{array}{rl}
\pG^2\,=\!\!&\!\!-\,4\big/\bigl(q^2+1\bigr)\,,\\[0.2em]
\pG\,=\!\!&\!\!p\,+\,(q^2+1)\dS_m\,,
\end{array}\! \right\} \hskip 5mm\Longleftrightarrow\hskip 5mm
\left\{\!\begin{array}{rl}
q^2+1\,=\!\!&\!\!-\,4/\pG^2\,,\\[0.2em]
p\,=\!\!&\!\!\pG\,+\,4\dS_m/\pG^2\,,
\end{array}\right.
\ee
where
\be\se
\dS_m\,=\,E(m\pi)^2.
\ee
\eme
The initial behaviour
\be
\label{initial-behaviour}
\chit_m\,\approx\,\dfrac{8t^{3/2}}{3\sqrt\pi}\,\dfrac{\IR_1\bigl(m\pi r\bigr)}{\IR_1\big(m\pi \ell\bigr)}\hskip 10mm (t\ll 1)
\ee
is recovered on expanding the integrand of the inverse-LT $\LC^{-1}_p\bigl\{\chith_m\bigr\}$  of $\chith_m$ defined by (\ref{pulse-m-LT}$a$) under the limit $p\to \infty$, for which $q\to \iR$ (see (\ref{pulse-m-LT}$c$)) and noting the results (\ref{soft-trigger}$a$,$b$). The initial response ($\propto t^{3/2}$) of (\ref{initial-behaviour}) is ``softer'' than the impulsive response \OSD{2.10$c$} to the $\EG$-trigger of Part~I.

For $t>0$ the LT-inversion of (\ref{pulse-m-LT}$a$) involves consideration of the contributions from various poles as well as the cuts at $p=\pm2\iR$. As in Part~I, we disregard the ageostrophic response linked to the poles $p=0$ and $\pG=0$, and restrict attention to the set $\daleth$ of poles $p=p_{mn}$, $p^*_{mn}$ (the superscript$^*$ denotes the complex conjugate)  identified by
\be
\label{q-j}
q\,=\,q_{mn}\,=\,{j_n}\bigl/({m\pi\ell})\,\,(>0)
\ee
\OSD{2.12$e$} determined by the real zeros $j_n$ of $\JR_1\big(m\pi q \ell\bigr)$. In turn, they define
\bme
\label{d-omega}
\begin{align}
\pG_{mn}\,=\,&\,\iR\omega_{mn}\,, &\, \omega_{mn}=\,&\,{2}\Big/\sqrt{q_{mn}^2+1}\,,\\
p_{mn}\,=\,&\,\iR\omega_{mn}\,-\,d_{mn}\,,&\, d_{mn}\,=\,&=\,4\dS_m\big/\omega_{mn}^2
\intertext{\OSD{2.12$d$}, \OSD{2.21$a$,$b$}. Other useful definitions are}
\FG_{mn}\,=\,&\,\HG^2_{mn} \big/2\,,  &    \HG_{mn}\,=\,&\,q_{mn}\omega_{mn}
\end{align}
\eme
\OSD{2.12$b$,$c$}. Our disregard of the ageostrophic response linked to the poles $p=0$ and $\pG=0$ has repercussions on the value that our solution exhibits at $r=\ell$. Essentially, instead of (\ref{pulse-m-bc}), it yields
\be
\label{pulse-m-bc-reduced}
r\chit_m\,=\, \ell\;\!\WG_{\tGH}(t) \hskip 12mm \mbox{at} \hskip 12mm r\,=\,\ell\,,
\ee
i.e., the quasi-steady response to the QG radial flow $-E^{-1/2}\ub_{\tQG}(\ell,t)$ part of the trigger (\ref{G-bc}) is thus omitted.

\subsection{The $r$-Fourier-Bessel series \label{r-Fourier-Bessel-series}}

In Part~I, we took advantage of the Fourier-Bessel series expansion for $\JR_1(m\pi q r)\big/\JR_1(m\pi q l)$ \OSD{B3} which permits us to express (\ref{pulse-m-LT}$a$) in the form
\bse
\label{pulse-m-LT-FBS}
\be
\left[\begin{array}{c}  \!\! \chith_m \!\!\\[0.3em]
    \!\!  \vth_m \!\!  \end{array}\right]\,
=\,\sum_{n=1}^\infty
\left[\begin{array}{c}  \!\! \chimrh_{mn} \!\!\\[0.3em]
    \!\!  \vmrh_{mn} \!\!  \end{array}\right]\,\dfrac{\JR_1(j_nr/\ell)}{j_n\JR_0(j_n)}
\hskip 10mm \mbox{on}  \hskip 10mm  0\le r < \ell\,,
\ee
where
\be
\left[\begin{array}{c}  \!\! \chimrh_{mn} \!\!\\[0.3em]
    \!\!  \vmrh_{mn} \!\!  \end{array}\right]\,=\,-\,\FG_{mn}\left[\begin{array}{c}  \!\! \pG \!\!\\[0.3em]
    \!\!  2m\pi \!\!  \end{array}\right]\,\dfrac{\pG\,\WGh(p)}{\pG^2+\omega_{mn}^2}\,.
\ee
\ese
An unfortunate feature of the Fourier-Bessel series expansion (\ref{pulse-m-LT-FBS}$a$) with inverse-LT
\bse
\label{pulse-m-LT-FBS-inverse}
\be
\left[\begin{array}{c}  \!\! \chit_m \!\!\\[0.2em]
    \!\!  \vt_m \!\!  \end{array}\right]\,
=\,\sum_{n=1}^\infty \left[\begin{array}{c}  \!\! \chimr_{mn} \!\!\\[0.2em]
    \!\!  \vmr_{mn} \!\!  \end{array}\right]\,\dfrac{\JR_1(j_nr/\ell)}{j_n\JR_0(j_n)}
 \hskip 10mm \mbox{on}  \hskip 10mm  0\le r < \ell\,,
\ee
where
\be
\left[\begin{array}{c}  \! \chimr_{mn} \!\!\\[0.2em]
    \!  \vmr_{mn} \!\!  \end{array}\right]\,=\,-\,\dfrac{\FG_{mn}}{2}\,\LC^{-1}_p\biggl\{
\biggl[\begin{array}{c}  \!\! \pG \!\!\\[0.2em]
    \!\!  2m\pi \!\!  \end{array}\biggr]\,\dfrac{\WGh(p)}{\pG-\iR\omega_{mn}} \biggr\}+\,\mbox{c.c.}
\ee
\ese
(``c.c.'' denotes complex conjugate), is that it necessarily  fails at $r=\ell$, because each eigenfunction $\JR_1(j_nr/\ell)$  vanishes there, $\JR_1(j_n)=0$. So it is not possible for (\ref{pulse-m-LT-FBS-inverse}$a$) to satisfy the reduced boundary condition $\chit_m(\ell,t)=\WG_\tGH(t)$ (\ref{pulse-m-bc}) except in the limiting sense $r\uparrow \ell$. 

A further reduction of the integral representation (\ref{pulse-m-LT-FBS-inverse}$b$) of $\vmr_{mn}$ is possible upon using the partial fraction decomposition 
\be
\label{p-f}
\dfrac{1}{\pG-\iR\omega_{mn}}\,=\,\dfrac{1}{\iR\omega_{mn}}\biggl[\dfrac{\pG}{\pG-\iR\omega_{mn}}\,-\,1\biggr].
\ee
The contribution to the inverse-LT integral $\LC^{-1}_p\bigl\{\vmrh_{mn}\bigr\}$ stemming from the second term $-1/(\iR\omega_{mn})$ is pure imaginary and so when added to its complex conjugate vanishes leaving
\vskip -3mm
\bse
\label{pulse-m-LT-FBS-W}
\be
\left[\begin{array}{c}  \!\! \chimr_{mn} \!\!\\[0.2em]
    \!\!  \vmr_{mn} \!\!  \end{array}\right]
\,=\,\dfrac12\left[\begin{array}{r}  \!\! -\,\FG_{mn}  \!\!\\[0.2em]
    \!\! \iR\, (j_n/\ell)\, \HG_{mn} \!\!  \end{array}\right]\WG_{mn}(t)+\,\mbox{c.c.}\,,
\ee
where $\WG_{mn}(t)$ has LT
\be
\WGh_{mn}(p)\,=\,\dfrac{\pG\WGh(p)}{\pG-\iR\omega_{mn}}\,.
\ee
The $\WG_{mn}$-notation is motivated by the property  $\WG_{mn}(t)=\WG(t)$, when $\omega_{mn}=0$. On setting $\WG(t)\,=\,\sumpm\bigl\{\WG^\pm(t)\bigr\}$ (\ref{our-trigger-b}$b$), we may usefully introduce the decomposition
\be
\WG_{mn}(t)\,=\,\sumpm\bigl\{\WG_{mn}^\pm(t)\bigr\}\,,
\ee
where $\WG_{mn}^\pm(t)$ has LT
\be
\WGh^\pm_{mn}(p)\,=\,\dfrac{\pG\WGh^\pm(p)}{\pG-\iR\omega_{mn}}\,=\,\dfrac{\pm\iR\,\pG}{p(p\pm 2\iR)^{1/2}(\pG-\iR\omega_{mn})}\,.
\ee
\ese

\subsection{The Laplace transform (LT-)inversion\label{LT-inversion}}

The LT-inversion of (\ref{pulse-m-LT-FBS-W}$b$) is awkward except for the limiting case $E=0$ studied in the next \S\ref{no-dis}. From a general point of view (i.e., $\forall E$), its constituents (\ref{pulse-m-LT-FBS-W}$c$) may be expressed as
\be
\label{pole-cut}
\WG^\pm_{mn}(t)\,=\,\WG^{\pm\daleth}_{mn}(t)\,+\,\WG^{\pm\leftrightarrows}_{mn}(t)
\ee
in terms of its pole $\daleth$ and cut $\leftrightarrows\,$-contributions to the inverse-LT $\LC^{-1}_p\bigl\{\WGh^\pm_{mn}(p)\bigr\}$.

\subsubsection{The pole $\daleth\,$-contribution\label{pole-section}}

On suitably modifying the development of {\S}I:$2.3$, subject to our approximation $Q=E^{1/2}\sigma=0$ (see (\ref{basic-approx}$a$)), the residues at the poles determine
\bse
\label{pole}
\be
\WG^{\pm\daleth}_{mn}(t)\,=\,\bigl(\CS^{\EG}_{mn}-\iR\SS^{\EG}_{mn}\bigr)\ES^{\RG\pm}_{mn}\,\exp\bigl[(\iR\omega_{mn}-d_{mn})t\bigr],
\ee
where
\begin{align}
\CS^{\EG}_{mn}-\iR\SS^{\EG}_{mn}\,=\,\biggl[\dfrac{\pG}{p}\od{p}{\pG}\biggr]_{\pG=\iR\omega_{mn}}\,=\,&\,\dfrac{\iR\omega_{mn}+2d_{mn}}{\iR\omega_{mn}-d_{mn}}\,,\\[0.1em]
\ES^{\RG\pm}_{mn}\,=\,\CS^{\RG\pm}_{mn}\pm\iR\SS^{\RG\pm}_{mn}\,\equiv\,&\,\dfrac{\pm \iR}{(p_{mn}\pm 2\iR)^{1/2}}\,.
\end{align}
\ese
Explicitly these coefficients are given by
\bme
\label{pole-coef}
\begin{align}
\hskip -5mm   \left[\begin{array}{l}  \!\!\CS^\EG_{mn}-1  \!\!\\[0.3em]
      \!\! \SS^\EG_{mn} \!\!  \end{array}\right]=\,&\,\dfrac{3d_{mn}}{\aleph^2_{mn}}
\left[\begin{array}{c}  \!\!-\,d_{mn}  \!\!\\[0.3em]
    \!\!\omega_{mn} \!\!  \end{array}\right],\hskip 5mm  &
\left[\begin{array}{c}  \!\! \CS^{\RG\pm}_{mn}   \!\!\\[0.3em]
    \!\!\SS^{\RG\pm}_{mn}  \!\!  \end{array}\right]=\,&\dfrac{1}{\sqrt{2}\,\bigl.\aleph^\pm_{mn}}
\left[\begin{array}{c}  \!\!\!\sqrt{\aleph^\pm_{mn}+d_{mn}}   \!\\[0.3em]
    \!\!\!\sqrt{\aleph^\pm_{mn}-d_{mn}}  \!  \end{array}\right],
\intertext{where}
\aleph_{mn}\,=\,&\,\sqrt{\omega_{mn}^2+d_{mn}^2}\,,& \aleph^\pm_{mn}\,=\,&\,\sqrt{(2\pm \omega_{mn})^2+d_{mn}^2}\,.
\end{align}
\eme
They determine
\bse
\label{pole-sol-coef}
\be
\WG_{mn}^\daleth(t)\,=\,\bigl(\CS^{\WG}_{mn}-\iR\SS^{\WG}_{mn}\bigr) \exp\bigl[(\iR\omega_{mn}-d_{mn})t\bigr]\,,
\ee
where
\be
\left[\begin{array}{c}  \!\!\CS^{\WG}_{mn}  \!\!\\[0.3em]
    \!\!\SS^{\WG}_{mn} \!\!  \end{array}\right]=\sumpm\Biggl\{
\left[\begin{array}{cc}  \!\! \CS^{\RG\pm}_{mn} &  \pm \SS^{\RG\pm}_{mn}\!\!\\[0.3em]
    \!\!\mp \SS^{\RG\pm}_{mn} & \CS^{\RG\pm}_{mn}\!\!  \end{array}\right]\Biggr\}
\left[\begin{array}{c}  \!\!\CS^{\EG}_{mn}  \!\!\\[0.3em]
    \!\!\SS^{\EG}_{mn} \!\!  \end{array}\right].
\ee
\ese
Finally the pole-part of the solution (\ref{pulse-m-LT-FBS-W}$a$), so determined, is
\bme
\label{pole-sol}
\be\se
\left[\begin{array}{c}  \!\! \chimr_{mn}^\daleth \!\!\\[0.4em]
    \!\!  \vmr_{mn}^\daleth \!\!  \end{array}\right]=\,-\,\left[\begin{array}{r}  \!\! \FG_{mn} \bigl(\CS^\WG_{mn}\cos\phi_{mn}+\SS^\WG_{mn}\sin\phi_{mn}\bigr)\!\! \\[0.4em]
    \!\! (j_n/\ell)\,\HG_{mn}\bigl(\CS^\WG_{mn}\sin\phi_{mn}-\SS^\WG_{mn}\cos\phi_{mn}\bigr)  \!\!  \end{array}\right]\exp\bigl(-\lambda_{mn}t\bigr)\,,
\ee
in which
\be
\phi_{mn}(t)\,=\,\omega_{mn}t\,, \hskip 20mm \lambda_{mn}\,=\,d_{mn}\,.
\ee
\eme
The corresponding $z$-Fourier series coefficients $[\,\chit_m^\daleth,\vt_m^\daleth\,]$ follow on substitution of (\ref{pole-sol}$a$) into (\ref{pulse-m-LT-FBS-inverse}$a$) (cf.~\OSD{2.23$a$,$b$}).

\subsubsection{The cut $\leftrightarrows\,$-contribution\label{cut-section}}

In addition to the poles, the integrand of inverse-LT integrals $\LC^{-1}_p\bigl\{\WGh^\pm_{mn}(p)\bigr\}$ possess cut-points at $p=\mp2\iR$. To determine the cut-contributions, we deform the contour of integration about them and consider the partial paths
\bse
\label{cut}
\be
p\,=\,p^\pm(s)\,=\,\mp2\iR-s^2 \hskip  15mm (-\infty<s<\infty)\,.
\ee
More precisely, each deformed partial contour is inwards below the cut $p=p^\pm(s)$  (from $s=-\infty$ to $s=0$ with $\Im\{s\}\uparrow 0$) returning outwards above (from $s=0$ to $s=\infty$ with $\Im\{s\}\downarrow 0$). The cut-contributions to the  inverse-LT integrals may then be expressed as
\be
\WG^{\pm\leftrightarrows}_{mn}(t)\,=\,\CG^{\pm}_{mn}(t)\exp(\mp\iR 2t)\,,
\ee
where
\be
\CG^{\pm}_{mn}(t)\,=\,-\,\dfrac{1}{\pi}\int_0^{\infty}\dfrac{2\pG^{\pm}\exp(- s^2 t)\,\dR s}{(2\mp\iR s^2)(\pG^{\pm}-\iR\omega_{mn})}\,,
\ee
in which $\pG^\pm$ solves
\be
p^\pm(s)\,=\,p(\pG^\pm)\,\equiv\,\pG^\pm\,+\,4\dS_m/(\pG^\pm)^2 
\ee
\ese
(see (\ref{pulse-m-LT}$c$)) with the root taken such that $\pG^\pm\to p^\pm\sim -s^2 $ as $s\to \infty$ and defined elsewhere by analytic continuation.

The proposed integration paths about the cuts $|\Im\{p\}|=2$ are distinct from the pole $\daleth$--locations $p=p_{mn}$, $p^*_{mn}$, which lie within the strip $|\Im\{p\}|<2$. This justifies our claim that the cut-contribution to the solution (\ref{pulse-m-LT-FBS-W}$a$) is
\be
\label{cut-sol}
\!\left[\begin{array}{c}  \!\! \chimr^{\leftrightarrows}_{mn} \!\!\\[0.3em]
    \!\!  \vmr^{\leftrightarrows}_{mn} \!\!  \end{array}\right]\,
=\,\dfrac12
\left[\begin{array}{r}  \!\! -\,\FG_{mn}\bigl(\CG^-_{mn}(t)+{\CG^+_{mn}}^{\!\!\!\!\!*}\;(t)\bigr)  \!\!\\[0.3em]
    \!\! \iR (j_n/\ell)\, \HG_{mn}\bigl(\CG^-_{mn}(t)-{\CG^+_{mn}}^{\!\!\!\!\!*}\;(t)\bigr) \!\!  \end{array}\right]\exp(2\iR t)+\,\mbox{c.c.}\,.
\ee
For Fourier $m$-modes with $m\ll E^{-1/2}$ implying $\dS_m\ll 1$ (see (\ref{pulse-m-LT}$d$)), we may solve (\ref{cut}$d$) iteratively to obtain
\be
\label{pG-expansion}
\pG^\pm\,=\,\mp 2\iR\biggl[1-\dfrac{16s^2}{(4+s^4)^2}\dS_m\biggr]\,-\,s^2+\dfrac{4(4-s^4)}{(4+s^4)^2}\dS_m+O(\dS_m^2), \hskip 10mm  \dS_m\ll 1\,.
\ee
Further, since $\dS_m\ll 1$ the cut-points $p=3\dS_{m}^{1/3}\exp(\iR 2\beta\pi/3)$ ($\beta=0,\,1,\,2$), at which $\dR p/\dR \pG=0$, are located close to $p=0$ and so do not introduce any complication with respect to the analytic continuation proposed to define the integrals (\ref{cut}$c$). In that limit, $\pG^\pm$ has the property 
\be
\label{pG-expansion-cc}
\pG^+\,=\,{\pG^-}^{*}  \hskip   20mm  \mbox{for $s$ real}
\ee
along the cuts upon which the integrals are taken. Reassuringly, Fourier $m$-modes with $m=O(E^{-1/2})$, for which $\dS_{m}=O(1)$, only exist on the Ekman layer length scale. That lies outside the range of applicability of our theory, and so such $m$-modes are irrelevant to us.

\subsubsection{An alternative direct LT-inversion\label{alternative-inversion}}

Though we have formulated our solution (\ref{pulse-m-LT-FBS-inverse}$a$) in terms of the Fourier-Bessel series \OSD{B3}, the original LT-formula (\ref{pulse-m-LT}$a$) may be investigated directly via its pole and cut-contributions. The pole-results are identical, while the cut-contribution, after the change of variable $q=\iR \rho$, is
\be
\label{cut-alternative}
\left[\begin{array}{c}  \!\! \chit_m^\leftrightarrows \!\!\\[0.2em]
    \!\!  \vt_m^\leftrightarrows \!\!  \end{array}\right]\,=\,-\,\dfrac{\exp(2\iR t)}{\pi}\int_{0}^{\infty}
\left[\begin{array}{c}  \!\! 1 \!\!\\[0.2em]
      \!\!  2m\pi/\pG \!\!  \end{array}\right]\,\dfrac{\IR_1\bigl(m\pi \rho r\bigr)}{\IR_1\big( m\pi \rho\ell\bigr)}\,\,\dfrac{\exp(-s^2t)}{1+\iR s^2/2}\,\,\dR s\,+\,\mbox{c.c.}\,.
\ee
The term given explicitly stems from the cut at $p=2\iR$ upon which $p=p^-=2\iR-s^2$ (the c.c.-term stems from the cut at $p=-2\iR$). On the cut, (\ref{pulse-m-LT}$b$-$d$) and (\ref{cut}$a$) determine
\bme
\label{cut-original}
\be
\rho^2\,=\,-q^2\,=\,\bigl(4+\pG^2\bigr)\big/\pG^2\,,  \hskip 20mm  \pG\,=\,2\iR-\sG^2\,,
\ee
where
\be\se
\sG^2\,=\,s^2\,-\,(1-\rho^2)\dS_m\,.
\ee
\eme

At the trigger location $r=\ell$, (\ref{cut-alternative}) gives
\bse
\label{cut-at-trigger} 
\begin{align}
\chit_m^\leftrightarrows(\ell,t)\,=\,&\,-\,\dfrac{\exp(2\iR t)}{\iR\pi}\int_0^\infty\dfrac{\exp(-qt)}{(q-2\iR)q^{1/2}}\,\,\dR q\,+\,\mbox{c.c.}\\
=\,&\,-\,\bigl((-\iR)^{1/2}\big/\sqrt{2}\bigr)
\erfc\bigl[(-\iR)^{1/2}\sqrt{2t}\bigr]\,+\,\mbox{c.c.}\nonumber\\
\,=\,&\,\WG(t)-1\,=\,\WG_\tGH(t)
\end{align}
\ese
(see (\ref{MF-towards-unified}$b$)). Here we have used eq.~(25) in \S4.2 of \cite{EMOT54I} to evaluate the integral in terms of $\erfc$ and used (\ref{our-trigger-a}$a$) and (\ref{our-trigger-b}$a$) to obtain the simple answer $\WG_\tGH(t)$, which confirms that $\chit_m^\leftrightarrows(r,t)$ meets the $\WG_{\tGH}$-trigger boundary condition (\ref{pulse-m-bc-reduced}).  This contrasts with the pole solution $\chit_m^\daleth$, generated from (\ref{pulse-m-LT-FBS-inverse}$a$) with $\chimr_{mn}^\daleth$ given by (\ref{pole-sol}), which simply vanishes at $r=\ell$, as argued in our discussion of (\ref{pulse-m-bc-reduced}).

When $E=0$, (\ref{cut-original}$c$) reduces to $\sG=s$ so that (\ref{cut-original}$a$,$b$)  simplify to
\bme
\label{cut-original-E=0} 
\be
\rho^2\,=\,\bigl(4+p^2\bigr)\big/p^2\,,  \hskip 20mm  \pG\,=\,p\,=\,2\iR-s^2\,,
\ee 
\eme
and hence the numerical evaluation of the integral (\ref{cut-alternative}) is straightforward. When $E\not=0$, the simplification (\ref{cut-original-E=0}) no longer applies. Instead we need the solution $\pG=\pG^{-}$ of the cubic $p(\pG^{-})=p^-=2\iR-s^2$ defined by (\ref{cut}$d$), which for $\dS_m\ll 1$ is given by (\ref{pG-expansion}). Since the Fourier-Bessel series representation, which we use for all our presented numerical results, vanishes at $r=\ell$, it only achieves the correct value $\chit_m(r,t)\to \WG_\tGH(t)$ in the limit $r\uparrow \ell$. To partially confirm that the discontinuity leads to no spurious behaviour in our numerical evaluation, we backed up our Fourier-Bessel series results by testing them against other results based on (\ref{cut-alternative}) by an approximate method outlined in appendix~\ref{Appendix}, which becomes exact for $E=0$. For the viscous problem, $E\not=0$, we need to make further approximations that
\begin{itemize}
\item[(i)] pertain to internal friction and
\item[(ii)] accommodate the Ekman boundary layers (indeed transient for inertial waves with frequency close to $2$) on $z=0$. 
\end{itemize}
\noindent
So the discontinuity at $r=\ell$, just mentioned, is only one issue amongst others that we discuss in \S\ref{dis}.

%%%%%%%%%%%%%%%%%%%%%%%%%%%%%%%%%%%%%
%%%%%%%%%%%%%%%%%%%%%%%%%%%%%%%%%%%%%
%%%%%       SECTION 3
%%%%%%%%%%%%%%%%%%%%%%%%%%%%%%%%%%%%%
%%%%%%%%%%%%%%%%%%%%%%%%%%%%%%%%%%%%%

\section{The inviscid limit, $E=0$, $d_{mn}=0$\label{no-dis}}

Though our governing equations (\ref{gov-eqs}) are formulated for $E$ finite, our boundary conditions (\ref{boundary-condits}) are only appropriate for an ideal fluid. Our strategy is to allow for internal viscous friction as encapsulated by (\ref{gov-eqs}$a$,$c$) and to capture the role of the boundary layers by judicious approximations that we discuss in the following section \S\ref{dis}. Here we focus on the unambiguous limit $E=0$. Of course, the spin-down problem is only meaningful for $E\not=0$, and so the results of this section must be interpreted in the sense of $E\downarrow 0$ outside vanishingly thin boundary layers. As we are interested in events on the spin-down time scale $E^{-1/2}$, it is important to appreciate that the $E=0$ results presented below are limited to $0<t\ll E^{-1}$. They do not apply on the longer diffusion time scale $E^{-1}$, over which the MF-shear layer, width $\Delta(t)=(Et)^{1/2}$ (\ref{Et-small}$b$), touching the $z=0$ rigid boundary expands to fill the entire layer. After that, the boundary condition (\ref{boundary-condits}$b$) at $r=\ell$ no longer applies, even in an approximate sense.

\subsection{The pole $\daleth\,$-contribution\label{pole-E0}}

On setting $E=0$ in (\ref{pulse-m-LT}$d$), by (\ref{d-omega}$d$) we have $d_{mn}=0$. Whence (\ref{pole}$b$,$c$) determine $\CS^{\EG}_{mn}-\iR\SS^{\EG}_{mn}=1$ and $\ES^{\RG \pm}_{mn}=\ES^{\RG 0\pm}_{mn}$, where
\bme
\label{pole-coef-compact}
\be
\ES^{\RG 0\pm}_{mn}\,=\CS^{\RG 0\pm}_{mn}\pm\iR\SS^{\RG 0\pm}_{mn}\,\equiv\,(\pm\iR)^{1/2}\Big/\sqrt{\aleph^{0\pm}_{mn}}\,, \hskip 15mm \aleph^{0\pm}_{mn}=2\pm\omega_{mn}\,.
\ee
In turn, (\ref{pole}$a$) reduces to
\be\se
\WG^{0\pm\daleth}_{mn}(t)\,=\,\ES^{\RG 0\pm}_{mn}\exp(\iR\omega_{mn}t)\,.
\ee
\eme
Further, the coefficients $\CS^\WG_{mn}$, $\SS^\WG_{mn}$ (\ref{pole-sol-coef}$b$), which define $\WG_{mn}^\daleth(t)$ (\ref{pole-sol-coef}$a$), reduce to
\bse
\label{pole-coef-0}
\be
\biggl[\begin{array}{c}  \!\!\CS^{\WG 0}_{mn} \!\!\\[0.2em]
    \!\!\SS^{\WG 0}_{mn} \!\!  \end{array}\biggr]=\,\sumpm\biggl\{\dfrac{1}{\sqrt{2\aleph^{0\pm}_{mn}}}
\biggl[\begin{array}{c}  \!\! 1 \!\!\\[0.2em]
    \!\! \mp 1 \!\!  \end{array}\biggr]\biggr\}=\,\dfrac{1}{\cos (2\alpha_{mn})}
\biggl[\begin{array}{c}  \!\! \cos \alpha_{mn} \!\!\\[0.2em]
    \!\! \sin \alpha_{mn}\!\!  \end{array}\biggr],
\ee
where
\be
\alpha_{mn}=\tan^{-1}\sqrt{\dfrac{2+\omega_{mn}}{2-\omega_{mn}}}\,\,-\,\,\dfrac{\pi}{4}\,,
\hskip 20mm 0<\alpha_{mn}<\dfrac{\pi}{4}\,.
\ee
\ese
They allow us to express pole-response (\ref{pole-sol}$a$) to the $\WG$-trigger in the compact form
\be
\label{pole-sol-0}
\!\biggl[\begin{array}{c}  \!\! \chimr_{mn}^{0\daleth} \!\!\\[0.2em]
    \!\!  \vmr_{mn}^{0\daleth} \!\!  \end{array}\biggr]=\,-\,\dfrac{1}{\cos(2\alpha_{mn})}\biggl[\begin{array}{r}  \!\! \FG_{mn}\,{\cos (\phi_{mn}-\alpha_{mn})}
 \!\! \\[0.2em]
    \!\! (j_n/\ell)\,\HG_{mn}\, {\sin (\phi_{mn}-\alpha_{mn})} \!\!  \end{array}\biggr]. 
\ee

The result (\ref{pole-sol-0}) differs from the pole-response to the $\EG$-trigger \OSD{2.22} (with $d_{mn}=Q=0$, as in \OSD{4.2}), through the presence of the non-zero phase angle $\alpha_{mn}$. For our $\WG$-trigger, the value $\alpha_{mn}=0$ only occurs in the  QG-limit $\omega_{mn}=0$. It happens when $q_{mn}=j_n/(m\pi\ell)\to\infty$, namely the short radial-$r$ length scale limit ($j_n\gg 1$). The alternative MF-limit $\omega_{mn}= 2$  corresponds to $\aleph^{0-}_{mn}\downarrow 0$ with
\be
\label{omega-MF}
\omega_{mn}\,=\,2m\pi\ell\big/\sqrt{j_n^2+(m\pi\ell)^2}\,\approx\, 2-q_{mn}^2 \hskip 10mm \mbox{for} \hskip 10mm q_{mn}\,=\,j_n/m\pi\ell\ll 1\,,
\ee
and is reached as $m\to\infty$, namely the short axial-$z$ length scale limit.

As $\omega_{mn}$ increases from $0$ to $2$, $\cos(2\alpha_{mn})$ decreases in concert from $1$ to $0$ (i.e., $\alpha_{mn}$ increases from $0$ to $\pi/4$). By implication our individual mode response to our $\WG$-trigger is greater than that for the $\EG$-trigger. This is most marked in the $q_{mn}\ll 1$ limit:
\bme
\label{omega-QG}
\be\te
\aleph^{0-}_{mn}\,\approx \,q_{mn}^2\,, \hskip 10mm \alpha_{mn}\,\approx\, \tfrac14\pi-\tfrac12q_{mn}\,,  \hskip 10mm \cos(2\alpha_{mn})\approx q_{mn}\,,
\ee
\be
{\FG_{mn}}\big/{\cos(2\alpha_{mn})}\,\approx \,2q_{mn}\,, \hskip 15mm  {\HG_{mn}}\big/{\cos(2\alpha_{mn})} \,\approx\, 2\,.
\ee
\eme
So despite the implied divergence of the coefficients $\CS^{\WG 0}_{mn}$ and  $\SS^{\WG 0}_{mn}$ (\ref{pole-coef-0}$a$) both $O(q_{mn}^{-1})$, the corresponding $\chit_m^{0\daleth}$-modes $O(q_{mn})$ have vanishing amplitude, while that of the $\vt_m^{0\daleth}$-modes remains bounded and $O(\ell^{-1})$ (see (\ref{pole-sol-0})).

\subsection{The cut $\leftrightarrows\,$-contribution \label{cut-E0}}

The cut-contributions $\chit_m^{0\leftrightarrows}$ and $\vt_m^{0\leftrightarrows}$ are less straightforward to calculate. It is possible to adopt the direct approach of \S\ref{alternative-inversion} and use (\ref{cut-alternative}) with (\ref{cut-original-E=0}). Here , however, we follow the more straightforward Fourier-Bessel approach of \S\ref{cut-section}. So as in \S\ref{pole-E0}, we consider the Fourier-Bessel series (\ref{pulse-m-LT-FBS-inverse}) with (\ref{pulse-m-LT-FBS-W}), where the $E=0$ values of coefficients $\CG^\pm_{mn}(t)$ (\ref{cut}$c$), which define $\WG^{\pm\leftrightarrows}_{mn}(t)$  (\ref{cut}$b$), reduce to
\bme
\label{cut-coef-0}
\se
\begin{align}
\CG^{0\pm}_{mn}(t)\,=\,&\,-\,\dfrac{2}{\pi}\int_0^{\infty}\dfrac{\exp(- s^2 t)}{\aleph^{0\pm}_{mn}\mp\iR s^2}\,\dR s\\
=\,&\,-\,\Bigl((\pm\iR)^{1/2}\Big/\sqrt{\aleph^{0\pm}_{mn}}\,\Bigr)\,\exp\bigl(\pm\iR\Phi^{\pm}_{mn}\bigr)\,\erfc\!\Bigl((\pm\iR)^{1/2}\sqrt{\Phi^\pm_{mn}}\,\Bigr)\,,
\end{align}
where
\be\te
\phi_{mn}(t)\,=\,\omega_{mn}t\,, \hskip 10mm  \Phi^\pm_{mn}(t)\,=\,\aleph^{0\pm}_{mn} t\,, \hskip 10mm  \Phi^\pm_{mn}(t)\mp\phi_{mn}(t)\,=\,2 t\,.
\ee
\eme
The result for $\CG^{0-}_{mn}(t)$ may be derived simply by the change of variables $t\mapsto  \aleph^{0-}_{mn}\Phi^-_{mn}/2$ in the related displayed integral (\ref{cut-at-trigger}$a$) (but see also (\ref{cut-alternative})). The corresponding form for $\CG^{0+}_{mn}(t)$ results from obvious minor changes in the derivation.

The merit of (\ref{cut-coef-0}$b$) is its sufficiently large $t$ asymptotic expansion
\be
\label{pulse-m-LT-FBS-W0-sym-cut}
\CG^{0\pm}_{mn}(t)\,\sim\,-\,\dfrac{1}{\aleph^{0\pm}_{mn}\sqrt{\pi t}}\sum_{k=0}^\infty\dfrac{1\cdot 3\cdots (2k-1)}{(\mp\iR)^{k}\bigl(2\Phi^\pm_{mn}\bigr)^k} \hskip 10mm \hbox{for} \hskip 10mm \Phi^{\pm}_{mn}\gg 1
\ee
(see (https://dlmf.nist.gov/7.12E1)), where  $1\!\cdot\! 3\cdots (2k-1)\equiv 1$ for $k=0$, but otherwise ($k\ge 1$) defined in the obvious way. On substitution of (\ref{pulse-m-LT-FBS-W0-sym-cut}) into (\ref{cut-sol}), we obtain
\bse
\label{sol-large-Phi}
\be
\left[\begin{array}{c}  \!\! \chimr^{0\leftrightarrows}_{mn} \!\!\\[0.3em]
    \!\!  \vmr^{0\leftrightarrows}_{mn} \!\!  \end{array}\right]
=\,\dfrac{1}{\sqrt{\pi t}}\,\sum_{k=0}^\infty
1\!\cdot\! 3\cdots (2k+1)\!\left[\begin{array}{r}  \!\! \FG_{mn}\,{\chixw}_{mnk}(t) \cos(2t-k\pi/2)\!\! \\[0.3em]
\!\! (j_n/\ell)\,\HG_{mn}\,{\vxw}_{mnk}(t)\sin(2t-k\pi/2)\!\! \end{array}\right],
\ee
where
\be
\biggl[\begin{array}{c}  \!\!\,\chixw_{mnk}(t) \!\!\\[0.1em]
    \!\!\vxw_{mnk}(t)  \!\!  \end{array}\biggr]
=\,\sumpm\biggl\{ \dfrac{1}{\aleph^{0 \pm}_{mn}(2\Phi^\pm_{mn}\bigr)^{k}}\biggl[\begin{array}{c}  \!\! 1 \!\!\\[0.2em]
    \!\! \mp 1 \!\!  \end{array}\biggr]\biggr\}\,.
\ee
\ese

The values of the $k=0$ coefficients defined by (\ref{sol-large-Phi}$b$) are 
\bse
\label{sol-large-Phi-k=0}
  \be
\biggl[\begin{array}{c}  \!\!\,\chixw_{mn0} \!\!\\[0.1em]
    \!\!\vxw_{mn0}  \!\!  \end{array}\biggr]
=\,\sumpm\biggl\{ \dfrac{1}{\aleph^{0 \pm}_{mn}}\left[\begin{array}{c}  \!\! 1 \!\!\\[0.2em]
    \!\! \mp 1 \!\!  \end{array}\right]\biggr\}=\,\dfrac{1}{\FG_{mn}}\left[\begin{array}{c}  \!\! 2 \!\!\\[0.2em]
    \!\! \omega_{mn} \!\!  \end{array}\right],
\ee
in which we have used the definitions (\ref{pole-coef-compact}$b$) and (\ref{d-omega}$b$,$e$,$f$). On substitution into (\ref{sol-large-Phi}$a$), the leading order $k=0$ terms yield
\be
\left[\begin{array}{c}  \!\! \chimr^{0 \leftrightarrows}_{mn} \!\!\\[0.3em]
    \!\!  \vmr^{0 \leftrightarrows}_{mn} \!\!  \end{array}\right]\,
\approx\,\dfrac{2}{\sqrt{\pi t}}\left[\begin{array}{r}  \!\!  \cos(2t) \!\!\\[0.3em]
    \!\! m\pi\, \sin(2t) \!\!  \end{array}\right],
\ee
\ese
where we have made further use of (\ref{q-j}). On noting that the limit $q\to 0$ of \OSD{B3} determines
\bse
\label{cut-sol-0}
\be
\dfrac{r}{2\ell}\,=\,-\,\sum_{n=1}^\infty\,\dfrac{\JR_1(j_nr/\ell)}{j_n\JR_0(j_n)}\,,
\ee
substitution of (\ref{sol-large-Phi-k=0}$b$) into (\ref{cut-sol}) and (\ref{pulse-m-LT-FBS-inverse}$a$) yields
\be
\left[\begin{array}{c}  \!\! \chit_m^{0\leftrightarrows} \!\!\\[0.2em]
    \!\!  \vt_m^{0\leftrightarrows} \!\!  \end{array}\right]\,
\approx\,-\,\dfrac{r}{\ell}\dfrac{1}{\sqrt{\pi t}}
\left[\begin{array}{c}  \!\! \cos(2t) \!\!\\[0.2em]
    \!\! m\pi\sin (2t)\,\!\!  \end{array}\right]\hskip 7mm \mbox{as}\hskip 5mm t\to\infty\,.
\ee
\ese
Finally substitution of (\ref{cut-sol-0}$b$) into (\ref{GH-FS}) yields
\be
\label{cut-sol-t-to-infinity-final}
E^{1/2}\left[\begin{array}{c}  \!\! \chi^{0\leftrightarrows} \!\!\\[0.2em]
    \!\!  v^{0\leftrightarrows} \!\!  \end{array}\right]\,\approx\,-\,\dfrac{E^{1/2}}{\sqrt{4\pi t}}\,\dfrac{r}{\ell}
\left[\begin{array}{c}  \!\! (z-1) \cos(2t) \!\!\\[0.2em]
    \!\!  \sin (2t)\!\!  \end{array}\right]\,=\,-\,
\left[\begin{array}{c}  \!\! \chib_{\tGH} \!\!\\[0.2em]
    \!\!  \vb_{\tGH} \!\!  \end{array}\right]\hskip 10mm (z>0)
\ee
(see (\ref{chib-MF}) and the scaling  (\ref{c-of-v})).

The conclusion, that the cut solution generated by (\ref{cut-sol}) with coefficients (\ref{cut-coef-0}$a$) tends to the asymptotic solution (\ref{cut-sol-t-to-infinity-final}) as $t\to\infty$, needs careful appraisal. To begin we note that, as $t\to\infty$, (\ref{cut-sol-t-to-infinity-final}) decays algebraically ($\propto t^{-1/2}$) and so is necessarily small compared to the pole contributions generated by (\ref{pole-sol-0}). That said, the final results (\ref{cut-sol-0}$b$) and (\ref{cut-sol-t-to-infinity-final}) hide the fact that for their validity, every $mn$-harmonic needs to have reached its asymptotic regime $t\gg 1/\aleph^{0 \pm}_{mn}$ (i.e., $\Phi^{\pm}_{mn}\gg 1$; see (\ref{cut-coef-0}$d$) and (\ref{pulse-m-LT-FBS-W0-sym-cut})). Indeed for $\omega_{mn}$ close to $2$, (\ref{omega-MF}) and (\ref{omega-QG}$a$) indicate that
\bse
\label{cut-sol-restrictions}
\be
1\big/\aleph^{- 0}_{mn}\,\approx\,(m\pi\ell/j_n)^2 \hskip 10mm \mbox{for} \hskip 10mm q_{mn}\ll 1 \hskip 3mm \Longleftrightarrow
\hskip 3mm 2-\omega_{mn}\ll 1\,.
\ee
So for any large fixed time $t$, the needed large $\Phi_{mn}^{-}$ is only achieved when
\be
m\,\ll \,(j_n/\pi\ell)\sqrt{t}\,,
\ee
\ese
which is impossible for all $m$. Indeed, even for the smallest $m=1$, the condition (\ref{cut-sol-restrictions}$b$) is only met for $t\gg (\pi\ell/j_n)^{2}$. As our numerical results are based on $\ell=10$, this asymptotic regime for the case $n=1$, namely $t\gg \ell^{2} =10^2$ is never reached.

Despite the above caveats, taken at face value, (\ref{cut-sol-t-to-infinity-final}) would suggest that the triggered flow $E^{1/2}\bigl[ \chi_m^{0\leftrightarrows}\,,\,v_m^{0\leftrightarrows}\bigr]$ might tend to cancel the trigger flow $\bigl[ \chib_{\tGH}\,,\,\vb_{\tGH}\bigr]$. If so, to effect that cancellation, a large-$\chi^{0\leftrightarrows}$ cell extending the full radial extent $0\le r<\ell$ needs to emerge. Indeed, on plotting  $\chi^{0\leftrightarrows}$ (not portrayed here), we found  that to be the case. Moreover, intriguingly on forming the sum $\chi^0=\chi^{0\daleth}+\chi^{0\leftrightarrows}$, the large extensive eddy suggested by (\ref{cut-sol-t-to-infinity-final}) evaporates, i.e., $\chi^{0\daleth}$ also exhibits an extensive cell that cancels it for all time. That finding, in itself, provides strong motivation for our study in the next \S\ref{combined-E0} of the combined trigger, which ought to automatically effect the cancellation.

\subsection{The combined-contribution $\forall t$\label{combined-E0}}

As the results of \S\ref{cut-E0} above indicate, we can never rely entirely on the large time asymptotics, for which the pole-cut decomposition (\ref{pole-cut}) is best suited. Instead, we now consider the entire $E=0$ LT-form (\ref{pulse-m-LT-FBS-W}$d$) for the Fourier-Bessel coefficients with inverse-LT
\bme
\label{pulse-m-LT-FBS-W0}
\se
\begin{align}
\WG^{0\pm}_{mn}(t)\,=\,&\,\LC^{-1}_p\biggl\{\dfrac{\pm\iR}{(p\pm 2\iR)^{1/2}(p-\iR\omega_{mn})}\biggr\} \\
=\,&\,\exp\bigl(\iR\omega_{mn}t\bigr)\,\LC^{-1}_q\biggl\{\dfrac{\pm\iR}{q(q\pm\iR\aleph^{0\pm}_{mn})^{1/2}}\biggr\} \hskip10mm (q=p-\iR\omega_{mn}) \\
=\,&\,\sqrt{2\big/\aleph^{0\pm}_{mn}}\;\exp\bigl(\iR\omega_{mn}t\bigr)\, \WG^{\pm}\bigl(\aleph^{0\pm}_{mn}t\big/2\bigr) \\
=\,&\Bigl((\pm\iR)^{1/2}\Big/\sqrt{\aleph^{0\pm}_{mn}}\,\Bigr)\,\exp\bigl(\iR\phi_{mn}\bigr)\,\erf\!\Bigl((\pm\iR)^{1/2}\sqrt{\Phi^\pm_{mn}}\,\Bigr)
\intertext{(use (\ref{our-trigger-a}$b$): $p\mapsto (2/\aleph^{0\pm}_{mn})q$ and (\ref{our-trigger-b}$a$): $t\mapsto (2/\aleph^{0\pm}_{mn})t/2$, $\WG^{0\pm}_{mn} \mapsto \sqrt{2/\aleph^{0\pm}}\WG^{0\pm}_{mn}$), which is the superposition}
\WG^{0\pm}_{mn}(t)\,=\,&\,\WG^{0\daleth\pm}_{mn}(t)\,+\,\CG^{0\pm}_{mn}(t)\exp(\mp\iR 2t)
\end{align}
\eme
of the pole (see (\ref{pole-coef-compact}$c$)) and cut (see (\ref{cut}$b$) with (\ref{cut-coef-0})) contributions. On use of (\ref{our-trigger-b}$a$) the form (\ref{pulse-m-LT-FBS-W0}$c$) may also be written
\be
\label{pulse-m-LT-FBS-W0-all-t}
\WG^{0\pm}_{mn}(t)\,=\,\sqrt{2\big/\aleph^{0\pm}_{mn}}\,\exp(\iR\phi_{mn})\Bigl[\SR\Bigl(\sqrt{{2}/{\pi}}\sqrt{\Phi^\pm_{mn}}\,\,\Bigr)\pm\iR \CR\Bigl(\sqrt{{2}/{\pi}}\sqrt{\Phi^\pm_{mn}}\,\Bigr)\Bigr].
\ee
Substitution into (\ref{pulse-m-LT-FBS-W}$a$,$c$) yields
\bse
\label{solution-E0}
\be
\left[\begin{array}{c}  \!\! \chimr^0_{mn} \!\!\\[0.3em]
    \!\!  \vmr^0_{mn} \!\!  \end{array}\right]=\,-\,
\left[\begin{array}{r}  \!\! \FG_{mn},\bigl(\CC^{\WG 0}_{mn}(t)\cos\phi_{mn} +\SC^{\WG 0}_{mn}(t)\sin\phi_{mn}\bigr)\!\!\\[0.3em]
    \!\! (j_n/\ell)\, \HG_{mn}\,\bigl(\CC^{\WG 0}_{mn}(t)\sin\phi_{mn} -\SC^{\WG 0}_{mn}(t)\cos\phi_{mn}\bigr)\!\!\end{array}\right],
\ee
where
  \be
\left[\begin{array}{c}  \!\! \CC^{\WG 0}_{mn}(t)   \!\!\\[0.3em]
    \!\!  \SC^{\WG 0}_{mn}(t)  \!\!  \end{array}\right]
=\,\sumpm\left\{\sqrt{\dfrac{2}{\aleph^{0\pm}_{mn}}}\,\left[\begin{array}{c}  \!\! \SR\bigl(\sqrt{{2}/{\pi}}\sqrt{\Phi^\pm_{mn}}\,\bigr)  \!\!\\[0.2em]
    \!\! \mp \CR\bigl(\sqrt{{2}/{\pi}}\sqrt{\Phi^\pm_{mn}}\,\bigr) \!\!  \end{array}\right]\right\}.
\ee
\ese
On sequential substitution of (\ref{solution-E0}$a$) into (\ref{pulse-m-LT-FBS-inverse}$a$) and (\ref{GH-FS}), they determine $[\chi,v]$.  Next, we describe limiting cases.

\subsubsection{The series solution\label{small_Phi}}

The entire function (\ref{pulse-m-LT-FBS-W0}$d$) has the expansion
\be
\label{W0-small-Phi}
\WG^{0\pm}_{mn}(t)\,=\,2\sqrt{\dfrac{t}{\pi}}\exp\bigl(\mp2\iR t\bigr)\sum_{k=0}^\infty\dfrac{(\pm\iR)^{k+1}\bigl(2\Phi^\pm_{mn}\bigr)^k}{1\cdot 3\cdots (2k+1)}
\ee
(see (http://dlmf.nist.gov/7.6.E2)), which is useful for $\Phi^\pm_{mn}\ll 1$. Substitution of (\ref{W0-small-Phi}) into (\ref{pulse-m-LT-FBS-W}$a$,$c$) yields
\bse
\label{sol-small-Phi}
\be
\left[\begin{array}{c}  \!\! \chimr^0_{mn} \!\!\\[0.2em]
    \!\!  \vmr^0_{mn} \!\!  \end{array}\right]
=\,2\sqrt{\dfrac{t}{\pi}}\,\sum_{k=0}^\infty
\dfrac{1}{1\!\cdot\! 3\cdots (2k+1)}\!\biggl[\begin{array}{r}  \!\!-\,\FG_{mn}\,\chixbr_{mnk}(t) \sin(2t-k\pi/2)\!\! \\[0.2em]
\!\! (j_n/\ell)\,\HG_{mn}\,\vxbr_{mnk}(t)\cos(2t-k\pi/2)\!\! \end{array}\biggr],
\ee
where
\be
\biggl[\begin{array}{c}  \!\!\,\chixbr_{mnk}(t) \!\!\\[0.1em]
    \!\!   \vxbr_{mnk}(t)  \!\!  \end{array}\biggr]
=\,\sumpm\biggl\{   (2\Phi^\pm_{mn}\bigr)^k   \biggl[\begin{array}{c}  \!\! 1 \!\!\\[0.2em]
    \!\! \pm 1 \!\!  \end{array}\biggr]\biggr\}\,.
\ee
\ese

When $t\ll 1$, the $k=0$ values $\chixbr_{mn0}=2$, $\,\chixbr_{mn1}=8t/3\,$ and  $\vxbr_{mn0}=0$, $\,\vxbr_{mn1}=4\omega_{mn}t/3\,$ determine the leading order approximation
\bse
\label{sol-small-t}
\be
\left[\begin{array}{c}  \!\! \chimr^0_{mn} \!\!\\[0.2em]
    \!\!  \vmr^0_{mn} \!\!  \end{array}\right]\,=\,-\,\dfrac{8t^{3/2}}{3\sqrt\pi}\,\FG_{mn}
\left[\begin{array}{c}  \!\! 1 \!\!\\[0.2em]
 \!\! 0  \!\!  \end{array}\right]  \,+\,O\bigl(t^{5/2}\bigr)\,.
\ee
With $q=\iR$ in \OSD{B$3$}, we have the identity
\be
\dfrac{\IR_1(m\pi r)}{\IR_1(m\pi\ell)}\,=\,-\,\sum_{n=1}^\infty\FG_{mn}\,\dfrac{\JR_1(j_nr/\ell)}{j_n\JR_0(j_n)}\,
\ee
\ese
which together with (\ref{pulse-m-LT-FBS-inverse}$a$) and (\ref{sol-small-t}$a$) recovers the initial behaviour (\ref{initial-behaviour}). This result, though reassuring, is of  lesser significance than the fact that, for $\omega_{mn}$ close to $2$, the $\Phi^-_{mn}(\ll 1)$ contributions to (\ref{sol-small-Phi}), and whence (\ref{sol-small-t}$a$) are useful for large $t$ in the range $1\ll t\ll 1/\aleph_{mn}^-$, while the  $\Phi^+_{mn}$ contributions must be determined on the basis of  $\Phi^+_{mn}=O(t)$ large, a limit we consider next.

\subsubsection{The asymptotic solution for $\Phi^\pm_{mn}\gg 1$\label{large_Phi}}

The appropriate apparatus for the case $\Phi^\pm_{mn}\gg1$, is encapsulated by pole-cut decomposition (\ref{pulse-m-LT-FBS-W0}$e$) and the discussion of its constituent parts $\WG^{0\daleth\pm}_{mn}(t)$ (see (\ref{pole-coef-compact}$c$) of \S\ref{pole-E0}) and $\CG^{0\pm}_{mn}(t)\exp(\mp\iR 2t)$ (see (\ref{cut-coef-0}$b$) of \S\ref{cut-E0}) respectively.  However, for completeness, we note that, when $\Phi^\pm_{mn}\gg1$, $\CR$ and $\SR$ have the leading order asymptotic forms
\be
\label{Fresnel-asym}
\Biggl[\begin{array}{c}  \!\! \CR\bigl(\sqrt{{2}/{\pi}}\sqrt{\Phi^\pm_{mn}}\,\,\bigr)  \!\!\\[0.2em]
    \!\!  \SR\bigl(\sqrt{{2}/{\pi}}\sqrt{\Phi^\pm_{mn}}\,\,\bigr) \!\!  \end{array}\Biggr]
\approx\,\Biggl[\begin{array}{c}  \!\! \tfrac12  \!\!\\[0.6em]
    \!\!  \tfrac12   \!\!  \end{array}\Biggr]+\,
\dfrac{1}{\sqrt{2\pi \Phi^\pm_{mn}}}\Biggl[\begin{array}{c}  \!\! \sin \Phi^\pm_{mn}  \!\!\\[0.5em]
    \!\!  -\, \cos \Phi^\pm_{mn}  \!\!  \end{array}\Biggr],
\ee
which upon substitution into (\ref{solution-E0}$b$) yield
\be
\label{pulse-m-LT-app-explicit-0-even-more-asym*}
\Biggl[\begin{array}{c}  \!\!\CC^{\WG 0}_{mn}(t)  \!\!\\[0.3em]
    \!\! \SC^{\WG 0}_{mn}(t)  \!\!  \end{array}\Biggr]
=\,\Biggl[\begin{array}{c}  \!\! \CS^{\WG 0}_{mn}  \!\!\\[0.3em]
    \!\! \SS^{\WG 0}_{mn}  \!\!  \end{array}\Biggr]\,-\dfrac{1}{\sqrt{\pi t}}\,\sumpm\left\{\dfrac{1}{\aleph^{0\pm}_{mn}}
\Biggl[\begin{array}{c}  \!\!  \cos \Phi^\pm_{mn} \!\!\\[0.3em]
    \!\! \pm\sin \Phi^\pm_{mn}  \!\!  \end{array}\right]\Biggr\}\,.
\ee
Substitution of the leading order terms $\CS^{\WG 0}_{mn}$ and $\SS^{\WG 0}_{mn}$, defined by (\ref{pole-coef-0}$a$), into (\ref{solution-E0}$a$) recovers the pole-contribution $\bigl[ \chit_m^{0\daleth}\,,\,\vt_m^{0\daleth}\bigr]$, namely the $E=0$ version of (\ref{pole-sol}). In addition, substitution of the following $O(t^{-1/2})$ terms of  (\ref{pulse-m-LT-app-explicit-0-even-more-asym*})  into (\ref{solution-E0}$a$), leads awkwardly, on use of (\ref{sol-large-Phi-k=0}$a$) and (\ref{cut-coef-0}$c$-$e$), to the leading order cut-contribution $\bigl[ \chit_m^{0\leftrightarrows}\,,\,\vt_m^{0\leftrightarrows}\bigr]$ (\ref{cut-sol-0}$b$). This route is circuitous and not recommended.

\subsection{Numerical results\label{Numerical-results}}

\begin{figure}
\centerline{}
\vskip 3mm
\centerline{
\includegraphics*[width=1.0 \textwidth]{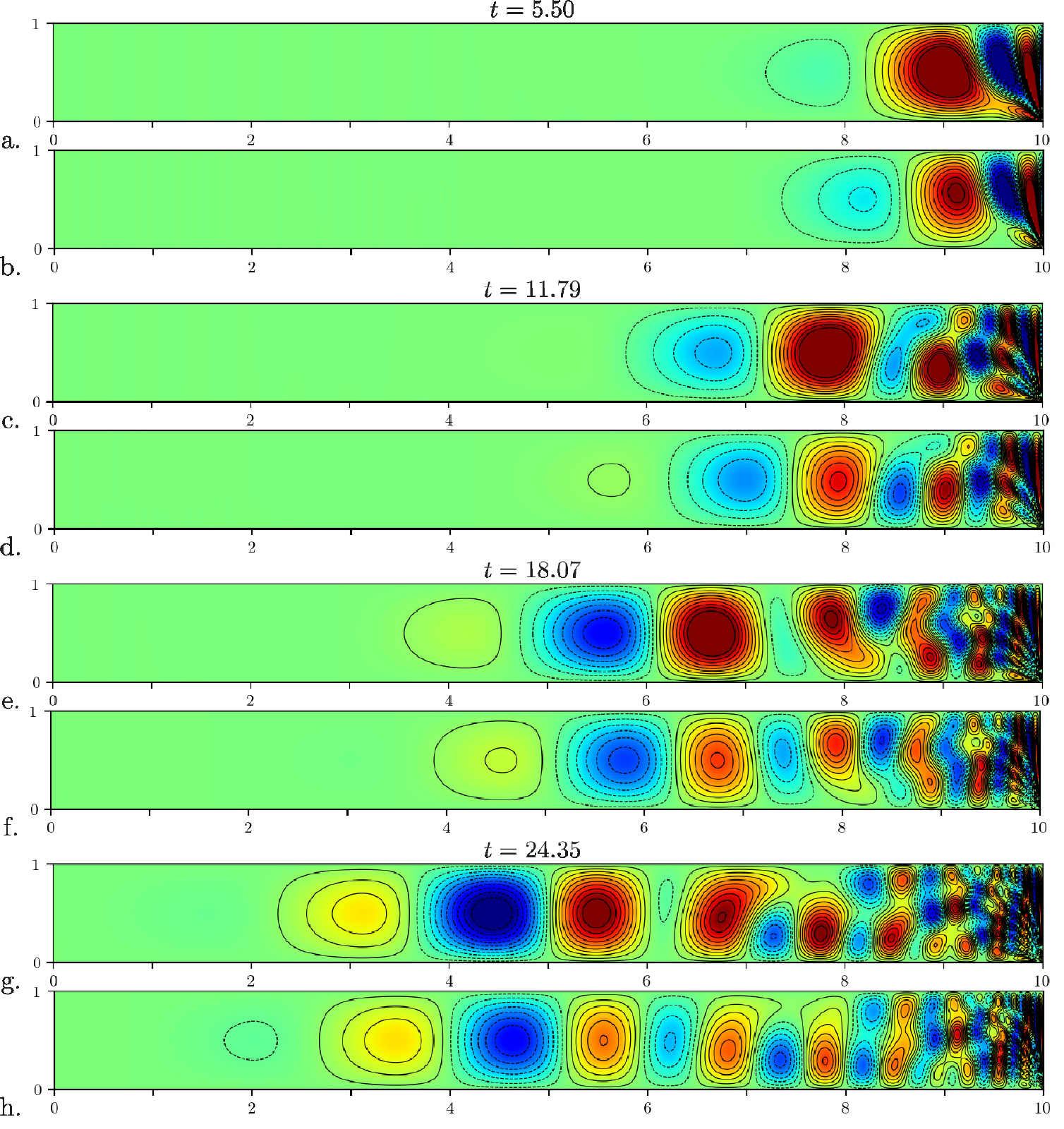}
}
\caption{(Colour online) $\chi^{\rm{wave}}$-contours (superscript$^{\rm{wave}}$ notation defined by (\ref{c-of-v}$a$)) in the $E\downarrow 0$ limit at the four instants $t=(N+\tfrac12)\pi/2$ ($N=3,\,7,\,11,\,15$), when $E^{-1/2}\chib_{\tGH}=0$: ($a$)-($b$), ($c$)-($d$), ($e$)--($f$), ($g$)--($h$) correspond to $t=5.50$, $11.79$, $18.07$, $24.35$ respectively. ($a$), ($c$), ($e$), ($g$) show $\chi^{\WG 0}$, resulting from our $\WG$-trigger (\ref{our-trigger-b}$b$); ($b$), ($d$), ($f$), ($h$) show $\chi^{\EG 0}$ determined by the $\EG$-trigger (\ref{basic-approx}$a$) (see figure~I:5).
}
\label{fig1-temp}
\end{figure}

\begin{figure}
\centerline{}
\vskip 3mm
\centerline{
\includegraphics*[width=1.0 \textwidth]{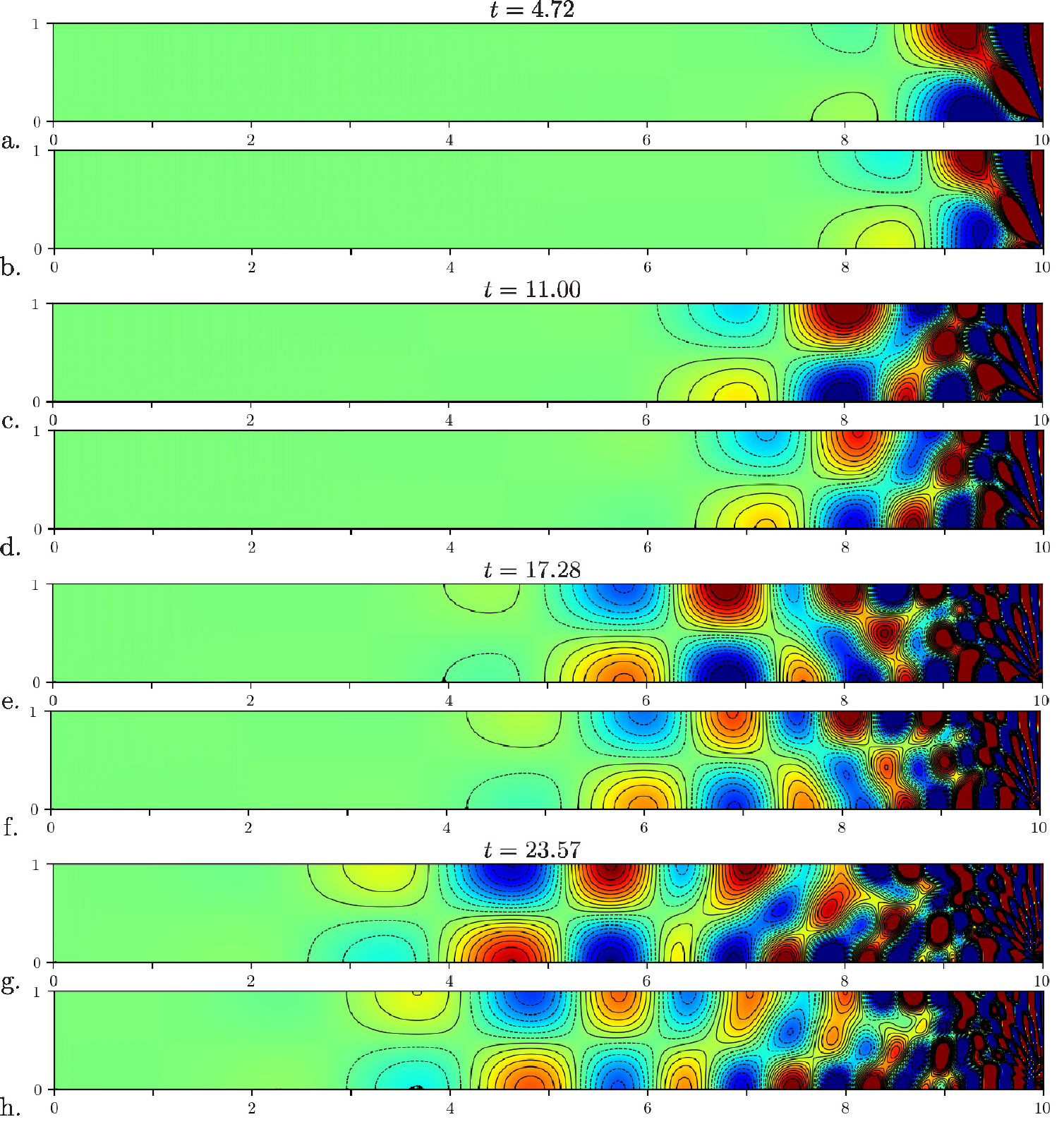}  
}
\caption{(Colour online) $v^{\rm{wave}}$-contours, as in figure~\ref{fig1-temp} but at the four instants $t=N\pi/2$ ($N=3,\,7,\,11,\,15$), when $E^{-1/2}\vb_{\tGH}=0$: ($a$)-($b$), ($c$)-($d$), ($e$)--($f$), ($g$)--($h$) correspond to $t=4.72$, $11.00$, $17.28$, $23.57$ respectively. ($a$), ($c$), ($e$), ($g$) show $v^{\WG 0}$; ($b$), ($d$), ($f$), ($h$) show $v^{\EG 0}$ (see figure~I:~6).
}
\label{fig2-temp}
\end{figure}

For the case $\ell=10$, we show in the alternate panels ($a$), ($c$), ($e$), ($g$) of  figures~\ref{fig1-temp}  and~\ref{fig2-temp} results for $\chi$ and $v$ respectively, which are obtained from (\ref{GH-FS}) on use of the Fourier-Bessel series (\ref{pulse-m-LT-FBS-inverse}$a$) with coefficients $\bigl[ \chimr_{mn}^{0}\,,\,\vmr_m^{0}\bigr]$ given by (\ref{solution-E0}). This straightforward approach raises issues of concern that we now address.

%%%%%%%%%%%%%%%%%%%%%%%%%%%%%%
%%%%%%%%%% FIGURE 1 %%%%%%%%%%
%%%%%%%%%%%%%%%%%%%%%%%%%%%%%%
%%%%%%%%%%%%%%%%%%%%%%%%%%%%%%
%%%%%%%%%%%%%%%%%%%%%%%%%%%%%%

%%%%%%%%%%%%%%%%%%%%%%%%%%%%%%
%%%%%%%%%% FIGURE 2 %%%%%%%%%%
%%%%%%%%%%%%%%%%%%%%%%%%%%%%%%
%%%%%%%%%%%%%%%%%%%%%%%%%%%%%%
%%%%%%%%%%%%%%%%%%%%%%%%%%%%%%

To begin we recall that the solution just described can be decomposed into pole and cut parts, whose coefficients can be traced to $\WG^{0\pm\daleth}_{mn}(t)$ and $\CG^{0\pm}_{mn}(t)\exp(\mp\iR 2t)$ respectively (see (\ref{pulse-m-LT-FBS-W0}$e$)). As we explained in \S\ref{alternative-inversion}, the primitive cut-solution  $\bigl[ \chit_m^\leftrightarrows\,,\,\vt_m^\leftrightarrows\bigr]$ given by (\ref{cut-alternative}) meets the required boundary condition $\chit_m^\leftrightarrows=\WG_\tGH(t)$ at $r=\ell$ (see (\ref{cut-at-trigger}$b$)). However, below (\ref{pulse-m-LT-FBS-inverse}$b$), we noted that each term of the Fourier-Bessel series  $\chit_m=\sum_{n=1}^\infty\chimr_{mn}\JR_1(j_nr\big/\ell)/\bigl(j_n\JR_0(j_n)\bigr)$  (see (\ref{pulse-m-LT-FBS-inverse}$a$)) vanishes at $r=\ell$, because there $\JR_1(j_nr/\ell)=\JR_1(j_n)=0$. So though $\chit^{0\leftrightarrows}_m=\sum_{n=1}^\infty\chimr^{0\leftrightarrows}_{mn}\JR_1(j_nr\big/\ell)/\bigl(j_n\JR_0(j_n)\bigr)$ correctly tends to $\WG_\tGH(t)$ as $r\uparrow \ell$, the vanishing of the sum at $r=\ell$, is of practical concern because convergence might be poor nearby. By contrast, the pole Fourier-Bessel series, that builds on $\chimr^{0\daleth}_{mn}$, vanishes at $r=\ell$, correctly so as explained below (\ref{cut-at-trigger}).

Since our theory is likely to work best for $t\gg 1$, there is a temptation to employ the large time asymptotics summarised in \S\ref{large_Phi} and simply retain the leading order term $O(t^{-1/2})$ of the cut-contribution $\bigl[ \chit_m^{0\leftrightarrows}\,,\,\vt_m^{0\leftrightarrows}\bigr]$, as approximated earlier by (\ref{cut-sol-0}$b$). That approach is unreliable because the Fourier-Bessel series coefficients $\bigl[ \chimr_{mn}^{0\leftrightarrows}\,,\,\vmr_{mn}^{0\leftrightarrows}\bigr]$ given by (\ref{sol-large-Phi-k=0}$b$) are only asymptotically correct when $m\ll (j_n/\pi\ell)\sqrt{t}$ (\ref{cut-sol-restrictions}$b$). From a slightly different perspective, when $2-\omega_{mn}$ is small, the requirement $\Phi^-_{mn}=\aleph^{0-}_{mn}t=(2-\omega_{mn})t \gg1$ for the validity of the asymptotic cut-values $\bigl[ \chimr_{mn}^{0\leftrightarrows}\,,\,\vmr_{mn}^{0\leftrightarrows}\bigr]$ (given to all orders by (\ref{sol-large-Phi})), provides a severe restriction, $t\gg 1/(2-\omega_{mn})$, on the time for their applicability. Indeed, for $m=1$, $n=1$, we have $\aleph^{0-}_{1,1}\approx q_{11}=(j_1/\pi\ell)^2=O\bigl(\ell^{-2}\bigr)$ for $\ell\gg 1$ (see (\ref{omega-MF})). As $\ell=10$ is adopted in our numerics, the results reported in figures~\ref{fig1-temp}  and~\ref{fig2-temp} never even reach $t=O\bigl(\ell^{2}\bigr)$, which is a minimal requirement for attaining a large time asymptotic regime for any of the harmonics.

Some interesting aspects of the solutions, already reported at the end of \S\ref{cut-E0}, are revealed by the pole-cut partition, when each part $[\chi^{0\daleth}\,,\,v^{0\daleth}]$ and $[\chi^{0\leftrightarrows}\,,\,v^{0\leftrightarrows}]$ is plotted separately (though not here). The most striking feature is the large $\chi^{0\leftrightarrows}$-cells that fill the container, just like the MF-trigger flow $\chib_{\tGH}$ (\ref{chib-MF}), which drives it. Being synchronised, the effect is most prominent at the times when $\chib_{\tGH}$ is maximised. Such extensive structure contrasts with the $\chi$-cells displayed in figure~\ref{fig1-temp}, which are restricted to a domain of limited extent inwards from the outer boundary  $r=\ell$. The large $t$ cut-approximation $\chi^{0\leftrightarrows}=(4\pi t)^{-1/2}(r/\ell)(z-1) \cos(2t)$ given by (\ref{cut-sol-t-to-infinity-final}) would account for such behaviour, but, as explained above, the approximation is unlikely to be valid at the moderately large times of interest to us. Despite these cautionary remarks, the $\chi^{0\daleth}$-plots also yield cells far from the outer boundary  with contour values of roughly the same magnitude but of opposite sign. This leads to the cancellation in the sum $\chi^0=\chi^{0\daleth}+\chi^{0\leftrightarrows}$, which is almost zero sufficiently close to the $r=0$ axis as in the plots on figure~\ref{fig1-temp}. Such cancellation is a feature of the small time  (rather  $\Phi^-_{mn}\ll 1$) series expansion of the combined solution $\bigl[ \chimr_{mn}^{0}\,,\,\vmr_m^{0}\bigr]$ given by (\ref{sol-small-Phi}) (see also $\WG^{0\pm}_{mn}(t)$ defined by (\ref{W0-small-Phi})).  This suggests that the only safe procedure is to use Fresnel integral form (\ref{solution-E0}) of the combined solution $\bigl[ \chit_m^{0}\,,\,\vt_m^{0}\bigr]$ valid for all time, as we have done.

The times adopted for the contour plots of $\chi$ in figure~\ref{fig1-temp} are limited to instants at which $\chib_{\tGH}\propto \sin(2t)$ vanishes ($\vb_{\tGH}\propto \cos(2t)$ maximised). By contrast in figure~\ref{fig2-temp}, the times are instants at which $\vb_{\tGH}$ vanishes ($\chib_{\tGH}$ maximised). These instants, taken to illustrate responses to our $\WG$-trigger (alternate panels ($a$), ($c$), ($e$), ($g$)), were chosen to coincide with those selected in Part~I to illustrate responses to the QG-trigger or simply $\EG$-trigger in figures~I:5 and I:6 and reproduced here (inter-spaced panels ($b$), ($d$), ($f$), ($h$)) for ease of comparison. In Part~I, we chose those instants to hide the MF-trigger flow in the full DNS, as it is only possible to isolate the triggered inertial waves in the DNS (or rather FNS, see \S\ref{numerics} below) at those instants. From another point of view, our decision not to provide plots of $\chi$ ($v$) at instants, when $\chib_{\tGH}$ ($\vb_{\tGH}$) is maximised and the cancellation of $\chi^{0\daleth}$ ($v^{0\daleth}$) with $\chi^{0\leftrightarrows}$ ($v^{0\leftrightarrows}$) is most pronounced, seems perverse. However, since the pulsating nature of each part at these times is not evident in the combined plots of $\chi$ ($v$), their omission is of no consequence.

In figures~\ref{fig1-temp}  and~\ref{fig2-temp}, it is striking to see how qualitatively similar the $\WG$-trigger response (alternate panels ($a$), ($c$), ($e$), ($g$)) is to the $\EG$-trigger response  (inter-spaced panels ($b$), ($d$), ($f$), ($h$)). The similarity reinforces our expectation that the $\EG$-trigger adopted in Part~I captures the essential mechanisms of inertial wave generation during the spin-down. Closer inspection reveals one significant distinction: In the case of the relatively large cells on the left (large $\ell-r$), the $\WG$-triggered cells are displaced to the left (decreasing $r$) relative to $\EG$-triggered cells. As explained in Part~I, the inertial waves propagate in the positive radial direction and so the $\WG$-triggered cells lag behind. This feature stems from the phase shifts $-\alpha_{mn}(<0)$ in the $\WG$-trigger pole-responses of all modes ($0<\omega_{mn}<2$) (identified in (\ref{pole-sol-0})) relative to the $\EG$-trigger modes with $\alpha_{mn}=0$.

Except for the above significant caveat, the cell structures are very similar. This is particularly true close to the right-hand boundary $r=\ell$, where the aforementioned phase shifts are less evident and the response is more sensitive to the current time trigger boundary condition. As time proceeds the $\WG$-trigger becomes ever closer to the $\EG$-trigger, i.e. $\WG(t) \to \EG(t)=1$ as $t\to\infty$ (see (\ref{basic-approx}$a$) and  (\ref{our-trigger-b}$b$)) with the consequence that their local ($\ell-r\ll 1$) responses become increasingly similar.

The true test of the merits of the more complicated $\WG$-trigger is whether or not its use improves the comparison with the full numerical results when finite $E$ effects are included. That we do in the following \S\S\ref{dis}~and~\ref{numerics}.

\newpage
%%%%%%%%%%%%%%%%%%%%%%%%%%%%%%%%%%%%%
%%%%%%%%%%%%%%%%%%%%%%%%%%%%%%%%%%%%%
%%%%%       SECTION 4
%%%%%%%%%%%%%%%%%%%%%%%%%%%%%%%%%%%%%
%%%%%%%%%%%%%%%%%%%%%%%%%%%%%%%%%%%%%

\section{Small dissipation, $E\ll 1$ \label{dis}}

The main motivation for considering the case $E\ll 1$ is to obtain formulae that may be used to compare with the full numerical results obtained for $E=10^{-3}$, $\ell=10$. A straightforward strategy, and one we indeed implement, is simply to apply the formulae of \S\ref{mathematical-problem} under the assumption $E\ll 1$. That was essentially the modus operandi of Part~I, where internal friction measured by $d_{mn}$ (\ref{d-omega}$d$) was retained and further Ekman layer damping measured by $d^\tE_{mn}$ ((\ref{EL-solution-parameters}$b$,$c$) below) was invoked. Our adoption of these dissipation concepts are summarised in \S\S\ref{Internal-friction} and \ref{EL-damping}.

Our objectives here are more ambitious than those of Part~I, for, on considering the more accurate $\WG$-trigger, we are aiming for results that more faithfully reproduce the full numerics, albeit external to all boundary layers. A key concern is signalled by the asymptotic result (\ref{cut-sol-t-to-infinity-final}) which indicates that, in the $E=0$ limit, the cut-contribution $\chit_m^{0\leftrightarrows}=\WG_\tGH$ at $r=\ell$ (\ref{cut-at-trigger}$b$) is finite, albeit decaying like $t^{-1/2}$. As discussed at length in \S\ref{Numerical-results}, this cut-feature is incompatible with the Fourier-Bessel series expansion, which is only valid for $r<\ell$. Essentially the Fourier-Bessel sum converges correctly to $\WG_\tGH(t)$ (as in (\ref{cut-at-trigger})) as  $r\uparrow\ell$, but not at $r=\ell$, where every harmonic vanishes. No such incompatibility arises for the pole-contribution. The obvious weakness of the Fourier-Bessel series for the cut-case is the spurious emphasis on small length scale modes in the vicinity of $r=\ell$, which will suffer considerable internal viscous dissipation. This may be of little consequence, as the region close to $r=\ell$ contains side-wall shear layers, where our analysis does not apply anyway. With that proviso, just as in the $E=0$ case, numerical results based on the primitive integral (\ref{cut-alternative}) for $\bigl[\chit_m^\leftrightarrows\,,\,\vt_m^\leftrightarrows\bigr]$ together with the definitions (\ref{cut-original}) would appear to be the safer strategy. We say ``safer'' as neither the Fourier-Bessel nor the primitive integral approach is perfect, owing our failure to implement robustly the consequences of the rigid boundary condition at $z=0$ on modes with frequency close to the MF-frequency 2, a matter that pertains particularly to the cut-contribution. In the light of these uncertainties, we obtained numerical results by both approaches. There were slight differences near $r=\ell$ reflecting their respective weaknesses. Since generally the entire Fourier-Bessel formulation (\ref{pulse-m-LT-FBS-inverse}$a$) for the combined sum gave results, which compared more favourably with the DNS, that is the method adopted here to generate our numerical results reported in \S\ref{numerics}. This approach also has the merit of being more straightforward to implement, with the nature of the approximations made (see \S\ref{approximation-appraisal}) more transparent.

\subsection{Internal friction\label{Internal-friction}}

Our strategy is to generalise the Fourier-Bessel method outlined in \S\ref{Fourier-series} for the poles to include the cut contribution as well. To that end, we replace (\ref{pole}$a$) by
\bse
\label{full-solution}
\be
\WG^{\pm}_{mn}(t)\,=\,\bigl(\CS^{\EG}_{mn}-\iR\SS^{\EG}_{mn}\bigr)\EC^{\RG\pm}_{mn}\exp\bigl[(\iR\omega_{mn}-d_{mn})t\bigr]\,,
\ee
in which $\ES^{\RG\pm}_{mn}$ (\ref{pole}$c$) has been replaced by
\be
\EC^{\RG\pm}_{mn}(t)\,=\,\CC^{\RG\pm}_{mn}(t)\pm\iR\SC^{\RG\pm}_{mn}(t)\,,
\ee
where
\be
\Biggl[\begin{array}{c}  \!\! \Bigl.  \CC^{\RG\pm}_{mn}(t) \Bigr. \!\!\\[0.2em]
    \!\!\Bigl. \SC^{\RG\pm}_{mn}(t) \Bigr. \!\!  \end{array}\Biggr]
=\,2\Biggl[\begin{array}{c}  \!\!  \CS^{\RG\pm}_{mn}\,\SR\bigl(\sqrt{{2}/{\pi}}\sqrt{\Phi^\pm_{mn}}\,\bigr)  \!\!\\[0.2em]
    \!\!  \SS^{\RG\pm}_{mn}\,\CR\bigl(\sqrt{{2}/{\pi}}\sqrt{\Phi^\pm_{mn}}\,\bigr) \!\!  \end{array}\Biggr].
\ee
\ese
This approach is guided by the following two limiting cases:
\begin{itemize}
\item[(i)] As $t\to\infty$,  both the Fresnel integrals in (\ref{full-solution}$c$) tend to $\tfrac12$ implying $\EC^{\RG\pm}_{mn}(t)\to\ES^{\RG\pm}_{mn}$, with the consequence that (\ref{full-solution}$a$) reduces to the pole result (\ref{pole}$a$), which provides the dominant part of the solution in the large $t$ limit. 
\item[(ii)] As $E\downarrow 0$, $d_{mn} \downarrow 0$,  we have $\CS^{\EG}_{mn}\to 1$, $\SS^{\EG}_{mn}\to 0$, while $2\CS^{\RG\pm}_{mn}$ and $2\SS^{\RG\pm}_{mn}$ both tend to $\sqrt{2\big/\aleph^{0\pm}_{mn}}$. With these limiting behaviours substituted into (\ref{full-solution}$b$,$c$), the formula  (\ref{full-solution}$a$) recovers the inviscid result (\ref{pulse-m-LT-FBS-W0-all-t}), i.e., $\WG^{\pm}_{mn}(t)\to\WG^{0 \pm}_{mn}(t)$.
\end{itemize}

Having made the anzatz (\ref{full-solution}), our pole-cut generalisation of (\ref{pole-sol-coef}) takes the form
\bse
\label{full-sol-coef}
\be
\WG_{mn}(t)\,=\,\bigl(\CC^{\WG}_{mn}(t)-\iR\SC^{\WG}_{mn}(t)\bigr) \exp\bigl[(\iR\omega_{mn}-d_{mn})t\bigr]\,,
\ee
\vskip-2mm
\noindent
where
\be
\Biggl[\begin{array}{c}  \!\! \CC^{\WG}_{mn}(t)  \!\!\\[0.3em]
    \!\! \SC^{\WG}_{mn}(t) \!\!  \end{array}\Biggr]=\sumpm\Biggl\{
\Biggl[\begin{array}{cc}  \!\! \CC^{\RG\pm}_{mn}(t) &  \pm \SC^{\RG\pm}_{mn}(t)\!\!\\[0.3em]
    \!\!\mp \SC^{\RG\pm}_{mn}(t) & \CC^{\RG\pm}_{mn}(t)\!\!  \end{array}\Biggr]\Biggr\}
\Biggl[\begin{array}{c}  \!\!\CS^{\EG}_{mn}  \!\!\\[0.3em]
    \!\!\SS^{\EG}_{mn} \!\!  \end{array}\Biggr],
\ee
\ese
while (\ref{pole-sol}$a$) becomes
\be
\label{full-sol}
\Biggl[\begin{array}{c}  \!\! \chimr_{mn} \!\!\\[0.4em]
    \!\!  \vmr_{mn} \!\!  \end{array}\Biggr]=\,-\,
\Biggl[\begin{array}{r}  \!\! \FG_{mn}\,\bigl(\CC^{\WG}_{mn}(t)\cos\phi_{mn} +\SC^{\WG}_{mn}(t)\sin\phi_{mn}\bigr)\!\!\\[0.4em]
    \!\! (j_n/\ell)\,\HG_{mn}\,\bigl(\CC^{\WG}_{mn}(t)\sin\phi_{mn} -\SC^{\WG}_{mn}(t)\cos\phi_{mn}\bigr)\!\!\end{array}\Biggr]
\,\exp(-\lambda_{mn}t)
\ee
with $\phi_{mn}(t)=\omega_{mn}t$ and $\lambda_{mn}=d_{mn}$ as in (\ref{pole-sol}$b$,$c$).

\subsection{Ekman layer damping\label{EL-damping}}

Ekman layer damping  modifications to the solution (\ref{full-sol}) are obtained by incrementing the frequency $\omega_{mn}$ and damping rate $d_{mn}$ to
\bme
\label{EL-solution-parameters}
\be
\phi_{mn}(t)\,=\,\bigl(\omega_{mn}+\omega^E_{mn}\bigr)t\,, \qquad\qquad \lambda_{mn}\,=\,d_{mn}+d^E_{mn}\,,
\ee
where
\se
\begin{align}  
d^E_{mn}\,=\,&\,\tfrac12 E^{1/2}\bigl(1-(\omega_{mn}/2)^2\bigr)^{1/2}\bigl[(1+\omega_{mn}/2)^{3/2}+(1-\omega_{mn}/2)^{3/2}\bigr]\,,\\
\omega^E_{mn}  \,=\,&\,\tfrac12 E^{1/2}\bigl(1-(\omega_{mn}/2)^2\bigr)^{1/2}\bigl[(1+\omega_{mn}/2)^{3/2}-(1-\omega_{mn}/2)^{3/2}\bigr]\,.
\end{align}
\eme
These formulae, respectively \OSD{2.25} and \OSD{2.24}, originate from the work of \cite{KB95} and \cite{ZL08}, as explained in \S{I:2.4}. There is an additional small correction $\epsilon^E_{mn}$ to the phase $\phi_{mn}(t)$ in (\ref{EL-solution-parameters}$a$), documented in \OSD{2.25$a$}. Its value, being small relative to the secular behaviour $\omega^E_{mn}t$, is ignored here, as in Part~I.

\subsection{An appraisal of the dissipation approximations\label{approximation-appraisal}}

The merit of the solution (\ref{GH-FS}) and (\ref{pulse-m-LT-FBS-inverse}$a$) utilising the approximate form (\ref{full-sol}) with the Ekman layer corrections (\ref{EL-solution-parameters}) is that, as $t\to \infty$, our damped ``wave'' response (in the sense of (\ref{c-of-v}) with the superscript `wave' dropped) $\chi^\tWG=\chi$, $v^\tWG=v$ to the $\WG$-trigger $-\ub_\tWG(\ell,t)=-\tfrac12 E^{1/2}\WG(t)$ (\ref{u-entire}$d$) tends to the damped ``wave'' response  $\chi^\tEG$, $v^\tEG$ to the $\EG$-trigger $-\ub_{\tQG}(\ell,t)= -\tfrac12 \sigma\kappa E^{1/2}\EG(t)$ (\ref{uQG-L}$a$), subject to the approximations $\kappa\sigma=1$, $\EG(t)=1$.

In Part~I, we ignored the $\EG$-trigger decay $\EG(t)=\exp(-E^{1/2}\sigma t)$, because it was found to have no influence on the numerical results, at least to graph plotting accuracy. That finding provided the motivation for our approximation $\EG(t)=1$ (\ref{basic-approx}) in our construction (\ref{our-trigger-a}) of the $\WG$-trigger (\ref{our-trigger-b}), in which any exponential decay has been ignored too. However, in Part~I, we retained the factor $\kappa\sigma=1+O(E^{1/2})$ in their definition (I:2.3) of the $\EG$-trigger. As our $\WG$-trigger is effectively based on $\kappa\sigma=1$, there are necessarily $O(E^{1/2})$ discrepancies. From that point of view, our retention of the actual values (\ref{pole-coef}$a$) of $\CS^\EG_{mn}=1+O(E)$ and $\SS^\EG_{mn}=O(E^{1/2})$ in the definition (\ref{full-sol-coef}$b$) of $\CC^{\WG}_{mn}(t)$ and $\SC^{\WG}_{mn}(t)$, rather than simply using  $\CS^\EG_{mn}=1$ and $\SS^\EG_{mn}=0$, is unnecessary. Since the cut-contribution decays as $t\to\infty$, we have retained the full definition of $\CS^\EG_{mn}$ and  $\SS^\EG_{mn}$ so that the persistent pole-contribution more faithfully reproduces the long time behaviour reported in Part~I (see also point (i) of \S\ref{Internal-friction} above).

Though most low order effects may be safely neglected, two apparently small ingredients, namely the frequency shift $\omega^E_{mn}$ and damping $d_{mn}+d^E_{mn}$ forms encapsulated by (\ref{EL-solution-parameters}$a$,$b$), must be retained,because of the secularities $\omega^E_{mn}t$ and $(d_{mn}+d^E_{mn})t$  linked to them. However, their implementation in (\ref{full-sol}), which builds on the non-inertial mode structures $\CC^{\RG\pm}_{mn}(t)$ and $\SC^{\RG\pm}_{mn}(t)$ (\ref{full-solution}$c$), can only be justified in the asymptotic limit $t\to\infty$ when the Fresnel integrals $S$ and $C$ both tend to $\tfrac12$ leaving the pure pole-contribution $\WG^{\pm\daleth}_{mn}(t)$ (\ref{pole}). That said, the internal friction damping $d_{mn}$ based on the mode shape may plausibly be reasonable for all time.

The notion of an oscillatory Ekman layer for $|\omega_{mn}|$ close to the MF-frequency $2$ needs careful assessment. To begin the boundary layer for each mode has a double layer structure, exhibiting widths
\bme
\label{wave-bl}
\be\se
\Delta^\pm_{mn}\,=\,\sqrt{2E\big/\aleph^{0\pm}_{mn}}
\ee
\citep[see, e.g.,][eq.~(2.8)]{KB95}. Essentially, the  Ekman layer, width
\be
\Delta^-_{mn}\,=\,\sqrt{Et_{mn}}\,,  \hskip 10mm\mbox{where} \hskip 10mm t_{mn}=(1-\omega_{mn}/2)^{-1}\,,
\ee
\eme
thickens indefinitely, as $t_{mn}\uparrow \infty$, so filling the entire layer as $|\omega_{mn}| \uparrow 2$. The prior transient evolution is characterised by an expanding viscous boundary layer width $\Delta(t)=\sqrt{Et}$
%(Et)^{1/2}
adjacent to $z=0$ similar to that identified by the MF-mode (\ref{MF-combined}). For $\Delta^-_{mn}\ll 1$, the final oscillatory steady state is reached when $\Delta(t)=\Delta^-_{mn}$ at time $t=t_{mn}$. For $t_{mn}\gg 1$, this may be longer than the times reached in our numerical investigations. So, when  $\Delta^-_{mn}=O(1)$  or $1\ll t \le O(t_{mn})$, the formulae (\ref{EL-solution-parameters}$c$,$d$) for $d^E_{mn}$ and $\omega^E_{mn}$ cease to be applicable. Nevertheless, since (\ref{EL-solution-parameters}$c$,$d$) predicts $d^E_{mn}\to 0$ and $\omega^E_{mn}\to 0$  as $|\omega_{mn}|\uparrow 2$, their use in that limit though inappropriate may well be harmless. The appearance of unjustifiable assumptions is a reminder that, owing to the omission of rigid boundary conditions in the set (\ref{boundary-condits}), we have not formulated a proper viscous problem. We therefore cannot analyse any boundary layer structures, albeit we attempt to retain the role of the Ekman jump condition. In the light of all these caveats, it is impossible to produce asymptotic results that are justifiable in all space or all time, when $0<E\ll 1$.

As a prelude to our discussion of numerical results in the following \S\ref{numerics}, we note that at the particular instants when $\chib_\tGH\approx 0$ and $\vb_\tGH\approx 0$ employed in figures~\ref{fig2} and~\ref{fig3}, the $\chi_\tIW(=E^{-1/2}\chi_\tGH+\chi)$ and $v_\tIW(=E^{-1/2}v_\tGH+v)$ (see (\ref{IW-W}) below) plots for $E=10^{-3}$ in panels ($b$), ($c$); ($e$), ($f$); ($h$), ($i$) approximate well  $\chi$ and $v$ for the same $E$. This fortuitous coincidence enables us to compare them with the corresponding $E=0$ results in figures~\ref{fig1-temp} and~\ref{fig2-temp}  panels ($a$), ($b$); ($c$), ($d$); ($e$), ($f$) respectively. The comparison of the $\WG$ and $\EG$-results, in the $E=10^{-3}$ case, appears to emphasise differences not so clearly evident in the $E=0$ case. Sufficiently far to the left (small $r$) the $\WG$ (and likewise the $\EG$) behaviours for $E=10^{-3}$ and $E=0$ are similar, albeit the $E=10^{-3}$ structures there, being of large scale, are only weakly damped. Sufficiently far to the right ($r$ close to $\ell$), the modes evident in the  $E=0$ case are predominantly short scale and heavily damped by internal friction in the $E=10^{-3}$ case. What little, that remains, shows considerable differences between the $\WG$ and $\EG$-results. One is tempted to conclude that our treatment of dissipation for the $\WG$-trigger maybe inadequate. Nevertheless, when we make appropriate comparisons with the Direct Numerical Simulation in the next \S\ref{numerics}, we reassuringly find that the $\WG$-trigger improves agreement considerably everywhere relative to that achieved by the $\EG$-trigger, so dispelling our fears of inadequacy.

%%%%%%%%%%%%%%%%%%%%%%%%%%%%%%%%%%%%%
%%%%%%%%%%%%%%%%%%%%%%%%%%%%%%%%%%%%%
%%%%%       SECTION 5
%%%%%%%%%%%%%%%%%%%%%%%%%%%%%%%%%%%%%
%%%%%%%%%%%%%%%%%%%%%%%%%%%%%%%%%%%%%
\newpage

\section{The $\WG$ and $\EG$-trigger predictions versus the filtered-DNS (FNS)\label{numerics}}

Results from the Direct Numerical Simulation (DNS) of the equations (\ref{gov-eqs}) governing the velocity $\vv_\tDNS$ subject to the complete set (no approximations) of initial \OSD{3.1} and boundary \OSD{3.2} conditions were described in \S{I}:3 and so will not be repeated here. From that solution of the properly posed viscous spin-down problem we removed the QG-part of the velocity to obtain (what we termed) the filtered-DNS, or simply the FNS-velocity, $\vv_\tFNS$. As there, we define its components by the recipe
\bme
\label{FNS}
\se
\begin{align}
v_\tFNS\,=&\,E^{-1/2}\bigl(v_{\tDNS}\,-\mu^{-1}\langle v_{\tDNS} \rangle\bigr)\,,\\
u_\tFNS\,=&\,E^{-1/2}u_{\tDNS}\,-\,\tfrac12(\sigma/\mu)\langle v_{\tDNS} \rangle
\intertext{\OSD{3.5}, \OSD{3.7$a$}, in which $\langle \bullet\rangle=\int_0^1\bullet\,\dR z$ is the $z$-average, and introduce}
%  d}
\chi_\tFNS\,=&\,E^{-1/2}\chi_{\tDNS}\,-\,\tfrac12(\sigma/\mu)(1-z)\langle v_{\tDNS} \rangle\, +\,O(E^{1/2})\,,
\end{align}
where
\be
\de
u_\tFNS\,=\,-\,\pd{\chi_\tFNS}{z}\,, \hskip 15mm w_\tFNS\,=\,\dfrac{1}{r}\pd{(r\chi_\tFNS)}{r}
\ee
\eme
\OSD{3.7$b$} with $\sigma=1+\tfrac34 E^{1/2}$  \OSD{1.18$e$} and $\mu=1-\tfrac12 E^{1/2}$ \OSD{1.19$c$}. Exterior to all boundary layers, the procedure removes the $O(1)$ azimuthal QG-velocity leaving only the small $O(E^{1/2})$ inertial wave part, which is why the FNS in (\ref{FNS}) is scaled up by a factor $O(E^{-1/2})$  relative to the DNS.

\subsection{The entire inertial waves (IW): $\vv_\tIW=E^{-1/2}\vv_\tGH+\vv^{\mathrm{wave}}$\label{IW-def}}

The inertial wave IW-velocity $E^{1/2}\vv_\tIW$ is composed of two parts:
\begin{itemize}
\item[(i)] The MF-waves $\vv_\tGH$ (see appendix I:A, but (\ref{MF-combined}) suffices for $t\gg1$);
\item[(ii)] the $\WG$-triggered waves $E^{1/2}\vv^{\mathrm{wave}}$ (see (\ref{c-of-v}$a$)).
\end{itemize}
(The superscript `wave' was omitted consistently throughout  \S\S\ref{mathematical-problem}--\ref{dis} but is reinstated here). Their combination is described by
\be
\label{IW-W}
\bigl[\,\chi_\tIW\,,\,v_\tIW\,\bigr]=\,E^{-1/2}\bigl[\,\chi_\tGH\,,\,v_\tGH\,\bigr]+\bigl[\,\chi^{\mathrm{wave}}\,,\,v^{\mathrm{wave}}\,\bigr],
\ee
as in \OSD{3.4} and \OSD{3.6$b$}. Our $\WG$-triggered waves are defined by the $z$-Fourier series (\ref{GH-FS}) and the $r$-Fourier-Bessel series (\ref{pulse-m-LT-FBS-inverse}$a$) utilising the approximate form (\ref{full-sol}) with the Ekman layer corrections (\ref{EL-solution-parameters}). It is important to appreciate at the outset that, whereas the expanding shear layer width $\Delta(t)=\sqrt{Et}$ captured at large $t$ by (\ref{MF-combined}) is retained in our MF-description (i), our procedures prohibit us from identifying any such comparable behaviour in the triggered inertial waves (ii) with frequency close to 2, a matter we return to in our final paragraph of the following \S\ref{comparison}.

Our main objective is to compare, in figures~\ref{fig1}--\ref{fig4}, the IW-response (\ref{IW-W}), identified by
\bse
\label{IW-WG-EG}
\begin{align}
\bigl[\,\chi_\tIW^\tWG\,,\,v_\tIW^\tWG\,\bigr],
\intertext{due to our $\WG$-trigger (\ref{u-entire}$d$),  in which $\WG(t)$ is defined by (\ref{our-trigger-b}$b$), with the IW-response} 
\bigl[\,\chi_\tIW^\tEG\,,\,v_\tIW^\tEG\,\bigr]
\end{align}
\ese
due to the $\EG$-trigger (\ref{uQG-L}$a$) previously reported in figures~I:1--4.

\subsection{Comparison of the FNS with the IW-results\label{comparison}}

\begin{figure}
\centerline{}
\vskip 3mm
\centerline{
\includegraphics*[width=1.0 \textwidth]{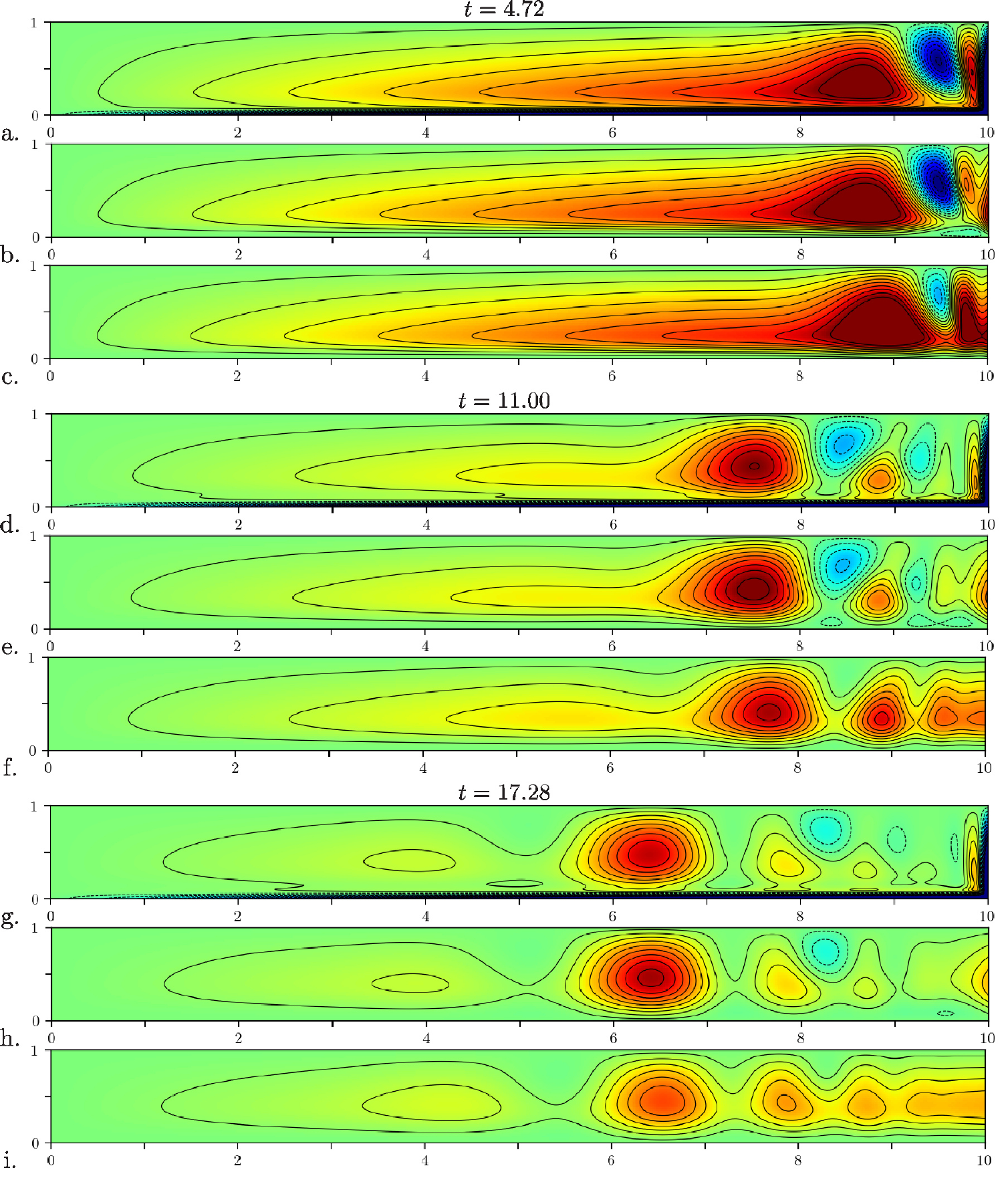}
}
\caption{(Colour online) The case $E=10^{-3}$, $\chi$-contours at three distinct instants $t=N\pi/2$ ($N=3,\,7,\,11$) when $E^{-1/2}\chib_{\tGH}$ is maximised: ($a$)--($c$), ($d$)--($f$), ($g$)--($i$) correspond to $t=4.72$, $11.00$, $17.28$ respectively.  ($a$), ($d$), ($g$) show $\chi_{\tFNS}$ (see figure~I:1($b$), ($e$), ($h$)); ($b$), ($e$), ($h$) show $\chi_\tIW^\tWG$; ($c$), ($f$), ($i$) show $\chi_\tIW^\tEG$ (see figure~I:~1($c$), ($f$), ($i$)) (colour scale from $-1$ to $1$).}
\label{fig1}
\end{figure}

\begin{figure}
\centerline{}
\vskip 3mm
\centerline{
\includegraphics*[width=1.0 \textwidth]{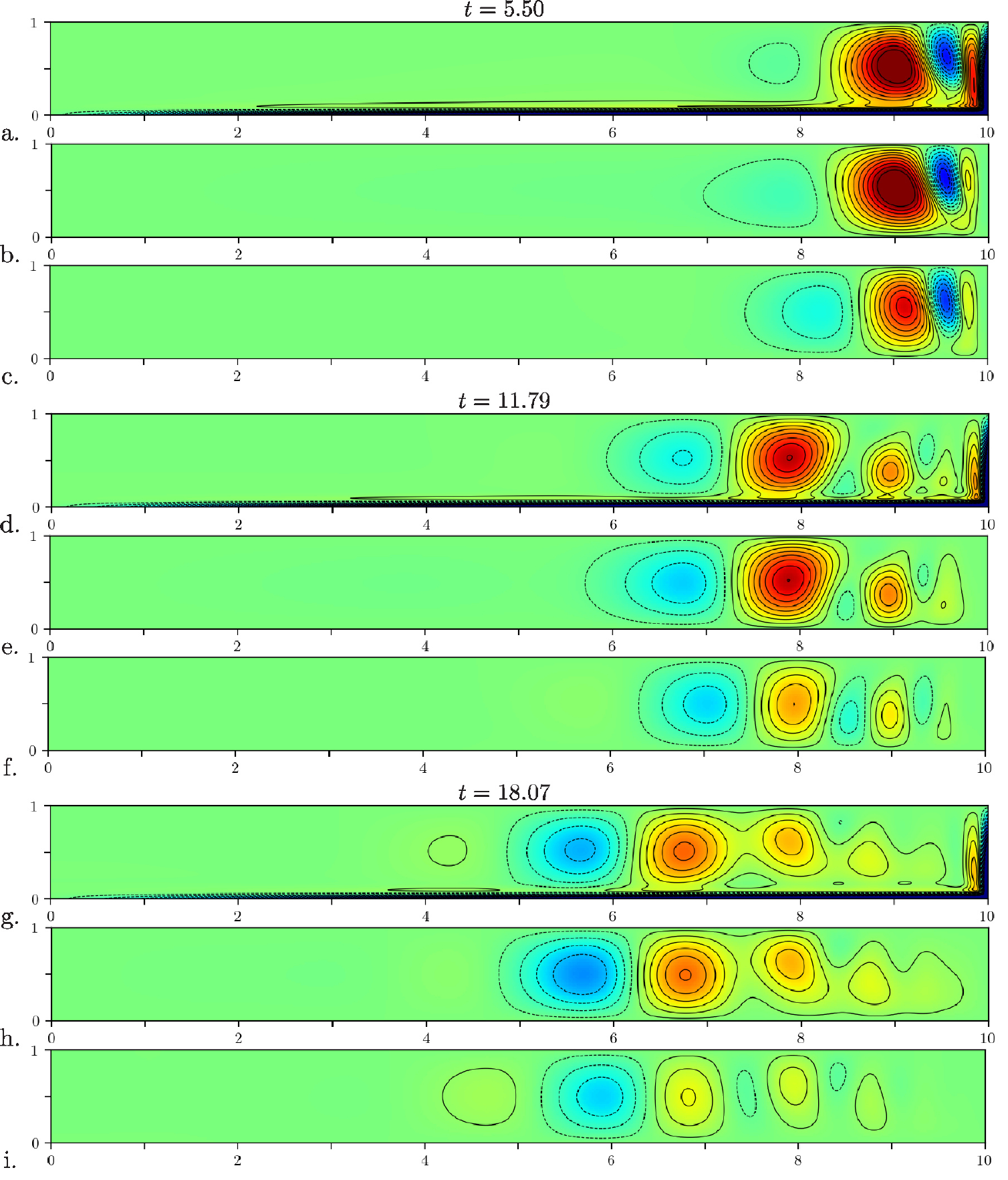}  
}
\caption{(Colour online) As in figure~\ref{fig1} but now at three distinct instants $t=(N+\tfrac12)\pi/2$ ($N=3,\,7,\,11$) at which $E^{-1/2}\chib_{\tGH}=0$. ($a$)--($c$), ($d$)--($f$), ($g$)--($i$) correspond to $t=5.50$, $11.79$, $18.07$ respectively.}
%       \hskip 50mm  -----------------------------------------------------------------------------------------------------------------------------
\label{fig2}
\end{figure}

\begin{figure}
\centerline{}
\vskip 3mm
\centerline{
\includegraphics*[width=1.0 \textwidth]{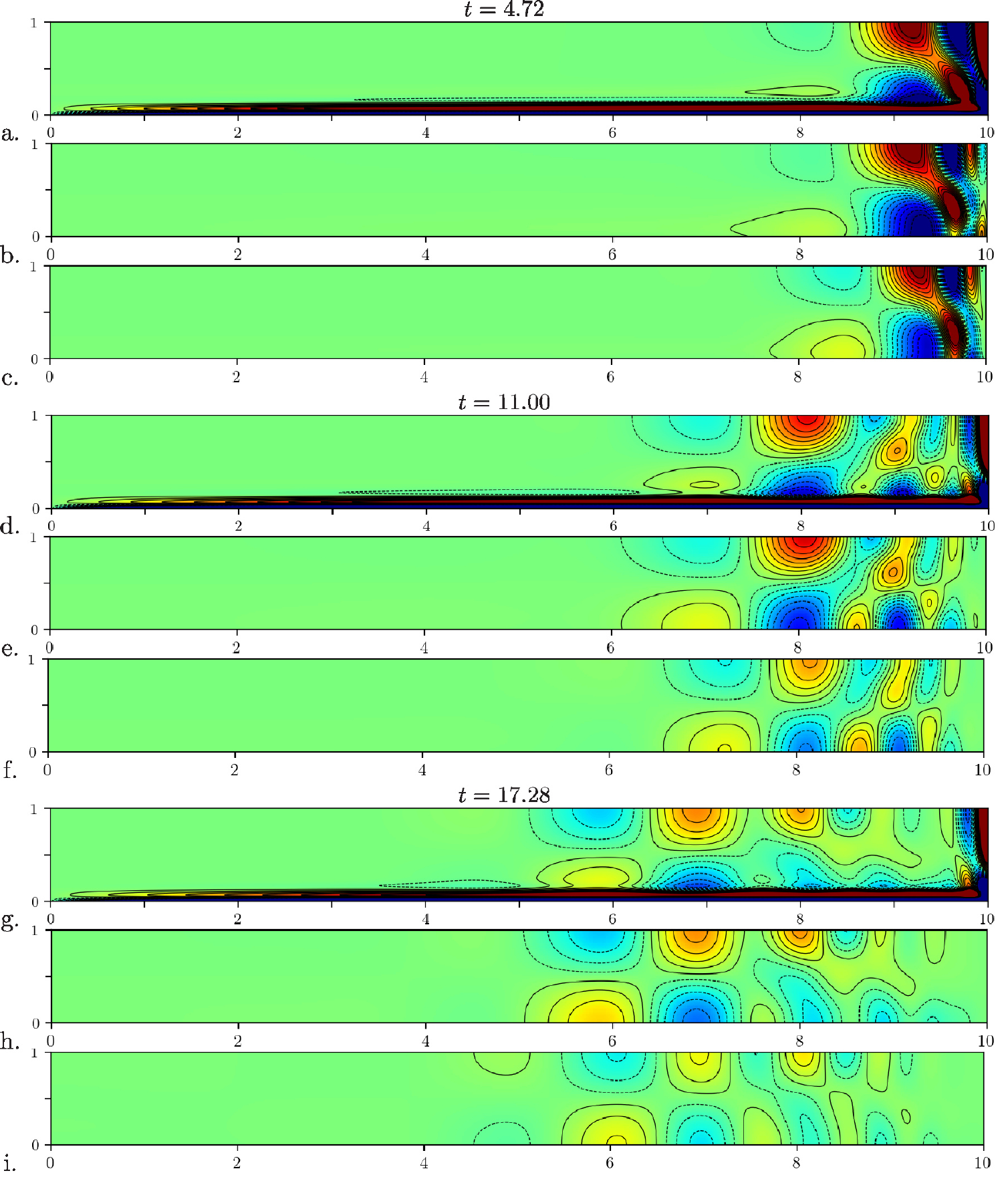}
}
\caption{(Colour online) As in figure~\ref{fig1} but now $v$-contours for the same instants, at which $E^{-1/2}\vb_{\tGH}=0$ (equivalent to $E^{-1/2}\chib_{\tGH}$ maximised). ($a$), ($d$), ($g$) show $v_{\tFNS}$ (see figure~I:3($b$), ($e$), ($h$)); ($b$), ($e$), ($h$) show $v_\tIW^\tWG$; ($c$), ($f$), ($i$) show $v_\tIW^\tEG$ (see figure~I:~3($c$), ($f$), ($i$)) (colour scale from $-5$ to $5$).}
\label{fig3}
\end{figure}

\begin{figure}
\centerline{}
\vskip 3mm
\centerline{
\includegraphics*[width=1.0 \textwidth]{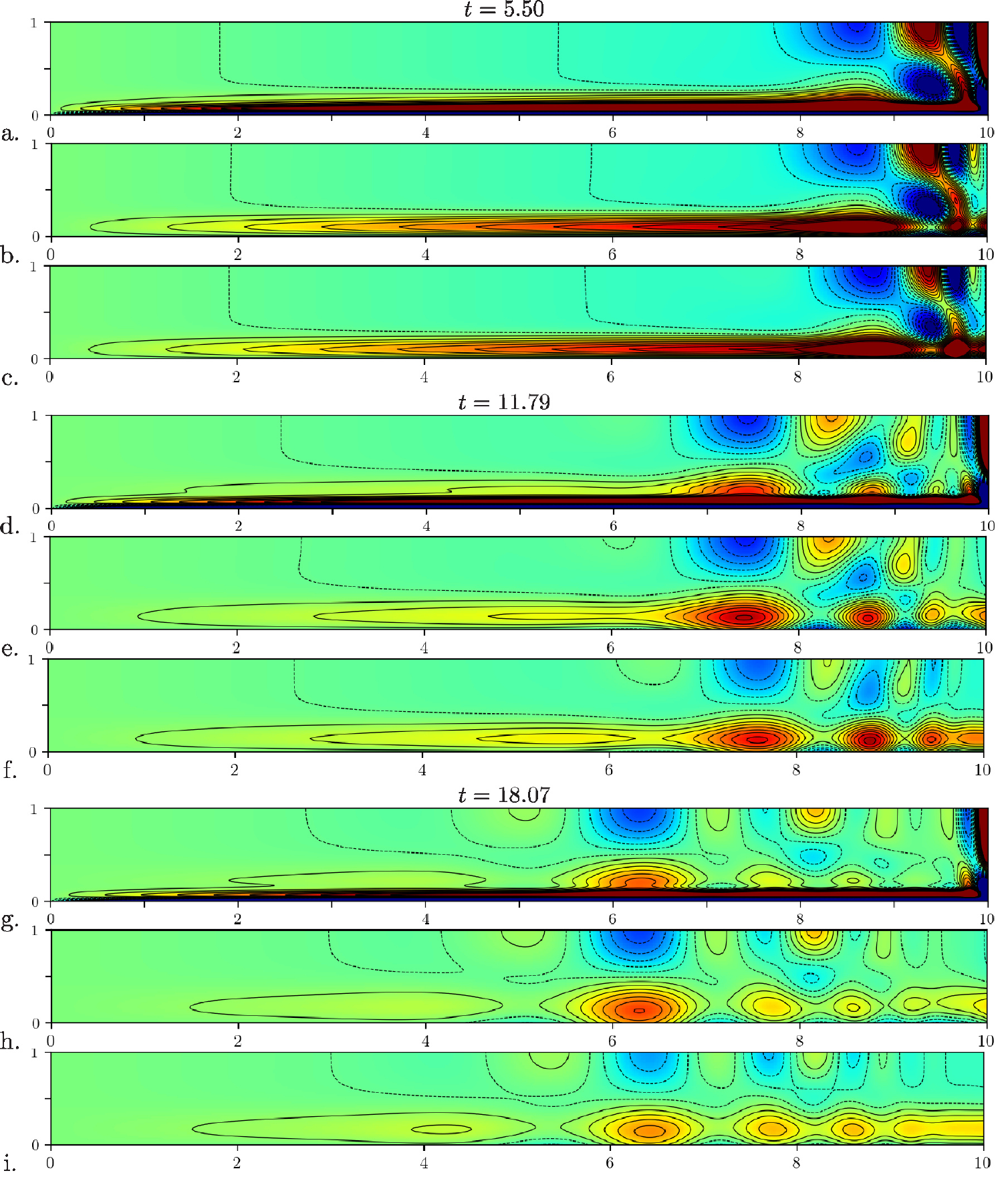}  
}          
\caption{(Colour online) As in figure~\ref{fig2} but now $v$-contours for the same instants, at which $E^{-1/2}\vb_{\tGH}$ is maximised (equivalent to $E^{-1/2}\chib_{\tGH}=0$). Panel description as in figure~\ref{fig3}.}
\label{fig4}
\end{figure}

The essential points of comparison between the FNS-results $[\chi_\tFNS\,,\,v_\tFNS]$ and the IW-results $[\chi_\tIW^\tEG\,,\,v_\tIW^\tEG]$ were explained in \S{I:3.2}. We summarise them here and identify the improvements made by  $[\chi_\tIW^\tWG\,,\,v_\tIW^\tWG]$. A key issue, already identified in \S\ref{Numerical-results}, is the choice of times for the plots. Since, for $t\gg 1$, $\chib_{\tGH}\propto \cos(2t)$ and $\vb_{\tGH}\propto \sin(2t)$ (see (\ref{chib-MF})), we note that $|\chib_{\tGH}|$ is maximised (figure~\ref{fig1}) and $\vb_{\tGH}\approx 0$ (figure~\ref{fig3}) at times $t=N\pi/2$, while  $\chib_{\tGH}\approx 0$ (figure~\ref{fig2}) and $|\vb_{\tGH}|$ maximised (figure~\ref{fig4}) at times $t=(N+\tfrac12)\pi/2$, where in both cases ($N=3,\,7,\,11,\,\cdots$).

In the case of the responses to the $\WG$-trigger, we made further checks. We compared $\chi^{\mathrm{wave}}$ with $\chi^\tWG_\tIW$ (figure~\ref{fig2}) at $t=(N+\tfrac12)\pi/2$ when $\chib_{\tGH}\approx 0$, as well as $v^{\mathrm{wave}}$ with $v^\tWG_\tIW$ (figure~\ref{fig3}) at $t=N\pi/2$ when $\vb_{\tGH}\approx 0$, and, not surprisingly, found them indistinguishable to graph plotting accuracy. At these instants further comparisons can be made of these figures with the triggered waves for $E=0$ illustrated in figures~\ref{fig1-temp} and~\ref{fig2-temp}, as explained in the last paragraph of \S\ref{approximation-appraisal}.  Interestingly, when the cell structures of $\chi^{\mathrm{wave}}$ and $v^{\mathrm{wave}}$ were plotted at the alternative times $t=N\pi/2$ and $(N+\tfrac12)\pi/2$ ($|\chib_{\tGH}|$ and $|\vb_{\tGH}|$ maximised) respectively, there was no essential change in their character from the plots at the aforementioned times $t=(N+\tfrac12)\pi/2$ and $N\pi/2$, for which well defined cells only extend a limited distance from the right-hand boundary $r=\ell$. This means that all the relatively intense structures exhibited by $\chi^\tWG_\tIW$ and $v^\tWG_\tIW$ on the left-hand side of figures~\ref{fig1} and \ref{fig4} stem from the MF-contributions $\chi_{\tGH}$ and $v_{\tGH}$ respectively. 

%%%%%%%%%%%%%%%%%%%%%%%%%%%%%%
%%%%%%%%%% FIGURE 3 %%%%%%%%%%
%%%%%%%%%%%%%%%%%%%%%%%%%%%%%%

%%%%%%%%%%%%%%%%%%%%%%%%%%%%%%
%%%%%%%%%%%%%%%%%%%%%%%%%%%%%%

%%%%%%%%%%%%%%%%%%%%%%%%%%%%%%
%%%%%%%%%% FIGURE 4 %%%%%%%%%%
%%%%%%%%%%%%%%%%%%%%%%%%%%%%%%

%%%%%%%%%%%%%%%%%%%%%%%%%%%%%%
%%%%%%%%%%%%%%%%%%%%%%%%%%%%%%

%%%%%%%%%%%%%%%%%%%%%%%%%%%%%%
%%%%%%%%%% FIGURE 5 %%%%%%%%%%
%%%%%%%%%%%%%%%%%%%%%%%%%%%%%%

%%%%%%%%%%%%%%%%%%%%%%%%%%%%%%
%%%%%%%%%%%%%%%%%%%%%%%%%%%%%%

%%%%%%%%%%%%%%%%%%%%%%%%%%%%%%
%%%%%%%%%% FIGURE 6 %%%%%%%%%%
%%%%%%%%%%%%%%%%%%%%%%%%%%%%%%

%%%%%%%%%%%%%%%%%%%%%%%%%%%%%%
%%%%%%%%%%%%%%%%%%%%%%%%%%%%%%

The various horizontal boundary layers adjacent to $z=0$, that appear on figures~\ref{fig1}--\ref{fig4} need careful identification. The Ekman layer, width $\Delta_E=E^{1/2}\approx 0.03$ for our choice $E=10^{-3}$, is associated with the relatively intense QG-flow. This Ekman layer is not filtered out from the DNS by the FNS and so is evident on the FNS-panels ($a$), ($d$), ($g$). Our opening remarks about the maximised MF-contributions focus our attention on two other important boundary layer matters.

Firstly, the MF-part $[\chi_\tGH,\,v_\tGH]$ of the IW-response (\ref{IW-W}) involves its thickening ($\Delta(t) =\sqrt{Et}$) MF-layer. This is visible in all panels of figure~\ref{fig1} ($|\chib_{\tGH}|$ maximised),  but is more forcefully illustrated by figure~\ref{fig4} ($|\vb_{\tGH}|$ maximised) on which the mainstream MF-flow $\vb_{\tGH}$ is identified by the vertical contours (at any rate to the left of the wave-cells). The MF-layer is corrupted on the right as the triggered wave-flow penetrates deeper to the left away from the outer boundary $r=\ell$. The elongated cells that emerge adjacent to the boundary $z=0$ are a blend of the wave-cells and the extensive MF-eddy that occupies the entire horizontal extent, $0<r<\ell$, of our cylinder.

Secondly, the triggered inertial waves possess Ekman boundary layers, whose consequences we incorporate (see \S\ref{EL-damping}). However, as we do not invoke their detailed analytic description, we are unable to visualise the layers themselves. As explained in \S\ref{approximation-appraisal}, those wave-modes with frequency $\omega_{mn}$ close to $2$ take a very long time for their Ekman layers to reach a steady state. For them steady Ekman layer theory does not apply. However, as their amplitude only increases linearly with time in the regime $\aleph_{mn}^\pm t \ll 1$ their boundary layer structure may be unimportant. Still, the essential point is that the expanding MF boundary layer issues  discussed above may also pertain to IW-modes with $2-\omega_{mn}\ll 1$. Without applying rigid boundary conditions explicitly, such transient features are outside the scope of our study. The  merit of our approximations is confirmed by very good agreement of the $\WG$-triggered responses with the realised FNS-results, as we now discuss.

So far we have mainly focused on the nature of the MF-trigger flow $E^{-1/2}[\,\chi_\tGH,\,v_\tGH]$, the triggered-modes $[\chi^{\mathrm{wave}},\,v^{\mathrm{wave}}\,]$ and the resulting IW-structure $[\chi_\tIW,\,v_\tIW]$ (see (\ref{IW-W})).  In figures~\ref{fig1}--\ref{fig4}, we now compare the FNS-solutions in panels ($a$), ($d$), ($g$) with our new $\WG$-trigger solutions in panels ($b$), ($e$), ($h$) and reproduce our Part~I $\EG$-trigger solutions in panels ($c$), ($f$), ($i$). Though the $\EG$-trigger solutions are qualitatively good,  the $\WG$-trigger solutions exhibit subtle but significant improvements upon which we now comment.

In the ante-penultimate paragraph of \S\ref{Numerical-results} we noted that, for the case $E=0$, the $\WG$-triggered cells are displaced to the left (decreasing r) relative to the $\EG$-triggered cells and explained the feature in terms of phase shifts $-\alpha_{mn}(<0)$ of all individual mode pole-responses. This effect leads to a remarkable improvement of the phase match by the $\WG$-triggered cells, which are now well synchronised with the FNS-cells, particularly on the left (sufficiently far from the outer $r=\ell$ boundary), where they are dominated the Fourier-$z$ series $m=1$ mode.

The increased intensity of the $\WG$-solution amplitude over the $\EG$-solution might simply reflect the improvement that ensues from use of the more accurate $\WG$-trigger. Alternatively, it might pertain instead more to the approximations made concerning dissipation. Be that as it may, it is remarkable how well the $\WG$-amplitudes agree with the FNS-amplitudes.

Particularly impressive is the improvement of detailed structure for moderate $\ell-r$ made by the $\WG$ over the $\EG$ solutions, a feature that was not so evident in the corresponding $E=0$ solution comparisons on figures~\ref{fig1-temp} and~\ref{fig2-temp}. This improvement is likely to be due to the fact that the $\WG$-trigger (\ref{our-trigger-b}$b$) is an almost perfect approximation of the early time trigger behaviour. Despite the superficial improvement, agreement is not perfect in the vicinity of $\ell-r\sim 1$. A likely explanation is that our internal friction anzatz in (\ref{full-solution}), which forms the basis of our assumed solution (\ref{full-sol}$a$), is only reliable for individual modes, when $\Phi^{-}_{mn}(t)\gg 1$. If the structure in this region is dominated by modes with $\Phi^{-}_{mn}(t)=O(1)$ or smaller, the assumed decay rate $d_{mn}$ possibly overestimates their dissipation. We add the caveat that as time proceeds there is an ageostrophic $E^{1/3}$-layer adjacent to the outer boundary $r=\ell$, that we cannot filter out and so pollutes the FNS-panels, when $\ell-r=O(0.1)$ for $E=10^{-3}$ or more likely a few multiples of $0.1$.

A slightly different perspective of the wave damping issues is suggested by the following comparisons. Inspection of figures~\ref{fig2} ($\chib_{\tGH}\approx 0$) and \ref{fig3} ($\vb_{\tGH}\approx 0$) shows tolerably good agreement between the FNS and $\WG$-triggered motions, which is possibly  accounted for by the absence of triggered modes with frequency close to 2. By contrast, figures~\ref{fig1} ($|\chib_{\tGH}|$ maximised) and \ref{fig4} ($|\vb_{\tGH}|$ maximised) show no agreement whatsoever for small $z$ within the expanding MF boundary layer. We have repeatedly emphasised our inability to reproduce such structures without applying the rigid boundary condition in a correct way, i.e., we do not address the fact that the triggered flow itself involves transient Ekman layers. This defect is compounded by the fact that the MF-trigger was approximated by $-\ub_{\tGH}(\ell,t)$ (\ref{MF-towards-unified}$a$), rather than $-u_{\tGH}(\ell,z,t)$, which means that we have totally ignored the boundary layer contribution $-u_{\tGH}(\ell,z,t)+\ub_{\tGH}(\ell,t)$ to the true trigger. From a more general point of view, this weakness is probably not as important as it first appears. Certainly as $\ell-r$ increases, owing to considerable wave interference, what remains has its origins in the early time nature of the $\WG$-trigger (our ``raison d'\^etre'' for use of  the term ``trigger''), which is well approximated by  $-\ub_{\tWG}(\ell,t)=-\ub_{\tGH}(\ell,t)-\tfrac12 E^{1/2}$ (see  (\ref{u-entire}$a$) and (\ref{MF-towards-unified}$a$,$b$)).

%%%%%%%%%%%%%%%%%%%%%%%%%%%%%%%%%%%%%
%%%%%%%%%%%%%%%%%%%%%%%%%%%%%%%%%%%%%
%%%%%       SECTION 6
%%%%%%%%%%%%%%%%%%%%%%%%%%%%%%%%%%%%%
%%%%%%%%%%%%%%%%%%%%%%%%%%%%%%%%%%%%%

\section{Concluding remarks\label{Discussion}}

The results presented here for the $\WG$-trigger, $-\ub_{\tWG}(\ell,t)=-\tfrac12 E^{1/2}\WG(t)$ (\ref{u-entire}$d$) together with the previous Part~I results for the (QG) $\EG$-trigger (\ref{uQG-L}$a$) provide a comprehensive description of the inertial waves that occur during the linear spin-down in a cylinder of large aspect ratio, $\ell\gg 1$. The partitioning of our complete study into two Parts~I and~II was guided by the following considerations:

The MF-waves identified by \cite{GH63} are transient and a manifestation of the transient Ekman layer in an unbounded cylinder ($\ell\to \infty$). For that reason we identified the $\EG$-trigger, associated with the persistent quasi-geostrophic spin-down, as the primary source of the additional inertial wave activity in the bounded cylinder ($\ell$~finite). That was sufficient reason for its study in Part~I. Moreover, being less complex than our $\WG$-problem, the $\EG$-problem is more amenable to detailed asymptotic analysis, in the $E\downarrow 0$ limit, well away from the axis ($\ell-r\ll \ell$), where the cylindrical geometry may be approximated as Cartesian, \S{I}:4.2. Accordingly, we were able to explain in \S{I}:5 the major inertial wave features, which include the fan-like structures emanating from the corner $(r,z)=(\ell,0)$ of ever decreasing length scale, and in \S{I}:6 the evolution of the large cells in the wave packet that moves to the left (negative $r$-direction); all visible in figures~\ref{fig1-temp} and~\ref{fig2-temp} for both the $\WG$ and $\EG$-trigger. However, to undertake such investigations for the $\WG$-trigger would be formidable and shed little new light on the physical mechanisms that operate. So detailed asymptotics similar to \S\S{I}:4--6 have not been attempted here.

The above considerations  might suggest that our new study of the $\WG$-trigger is unimportant. That overlooks the significant fact that during the early (rotation) time, $t=O(1)$, the MF-contribution $-\ub_{\tGH}(\ell,t)=-\tfrac12 E^{1/2} \WG_{\tGH}(t)$ is of comparable size to the QG-part $-\ub_{\tQG}(\ell,t)=-\tfrac12 E^{1/2}$, which results on making the approximations $\sigma\kappa=1$  and $\EG(t)=1$ in (\ref{uQG-L}$a$). Much of the later persistent wave response stems from the nature of that early time (and thus appropriately named) ``trigger''. Accordingly, a proper asymptotic solution of the spin-down for finite $\ell$ (large) must take account of the actual $\WG$-trigger based on the \cite{GH63} mainstream solution (their eq.~(3.17)), as interpreted by us in (\ref{u-entire}). On the one hand, in Part~I, by adopting the $\EG$-trigger, dominant for $1\ll t\ll E^{-1}$, we identified the basic mechanisms and produced results, which agreed surprisingly well with the FNS-results derived by filtering the DNS. On the other hand, here by use of the $\WG$-trigger, which is uniformly valid over the entire time interval  $0< t \ll E^{-1}$ including the crucial spin-down time $t=O\bigl(E^{-1/2}\bigr)$, we are able to identify significant improvements in the detailed structure. They are highlighted by the comparisons made in figures~\ref{fig1}-\ref{fig4} for the case $E=10^{-3}$, $\ell=10$ on the spin-down time $t\sim E^{-1/2}=10^{3/2}\sim 30$, over which the MF-boundary layer of width $\Delta(t)=\sqrt{Et}$ adjacent to $z=0$, remains thin ($\Delta(E^{-1/2})=E^{1/4}=10^{-3/4}\sim 0.2$).

The origin of the aforementioned finite $E$ improvements is elucidated by a comparison of the $\WG$ and $\EG$-triggered waves in the $E\downarrow 0$ limit  in the respective alternate panels ($a$), ($c$), ($e$), ($g$) and inter-spaced panels ($b$), ($d$), ($f$), ($h$) of figures~\ref{fig1-temp},~\ref{fig2-temp}. The time span encompassed by all our figures~\ref{fig1-temp}-\ref{fig4} is terminated, as in Part~I, at an appropriate instant before the wave activity has reached the axis $r=0$; a time span that increases with $\ell$. For after that, waves reflected at (or perhaps better crossing) the axis of symmetry lead to a confused picture that sheds no new light on the fundamental mechanisms identified in \S\S{I}:4--6.

As we explained in the ``Concluding remarks'' of Part~I, there has been a considerable amount of research on spin-up/down (see, e.g., \citealt{Letal12}, and references therein; from an overall perspective see \citealt{ZL17}). Particularly relevant to our studies here and in Part~I are those of \cite{KB95} and \cite{ZL08} for a circular cylinder with $\ell=O(1)$. They identified the free modes together with their decay rates. They did not address the matter of relative wave amplitude between individual modes during the spin-down process, nor for that matter their accumulated structure.  By that we mean that, like \citet{G68} before, they considered a model expansion of the combined $z$-Fourier (\ref{GH-FS}) and $r$-Fourier-Bessel (\ref{pulse-m-LT-FBS-inverse}$a$) series  type, but unlike in (\ref{pulse-m-LT-FBS-inverse}$a$) the individual mode amplitudes remained undetermined. This comparison highlights a technical matter. On the one hand, each mode in the studies of Kerswell \& Barenghi and Zhang \& Liao had a well defined complex exponential behaviour associated with the poles of a LT-solution.  On the other, our LT-solution has cut contributions, already present in our $\WG$-trigger LT-(\ref{our-trigger-a}) based on the transient unbounded mainstream flow defined by Greenspan \&~Howard, their eq.~(3.17). This asymptotic description of the flow, valid as $E\downarrow 0$, leads to all the difficulties that we encountered in the $0<E\ll 1$ context of \S\ref{dis} concerning how to perturb the Fresnel integral description appropriate to the limiting case $E=0$ discussed in \S\ref{no-dis}.  These cuts do not exist in the exact LT-solution eq.~(3.5) of Greenspan \&~Howard, as explained in their subsequent discussion. For that LT-solution they identify the approximate location of the poles in their eq.~(3.8) which are solely responsible for the transient solution, just as in the general approach of \citet{G68}, his eqs.~(2.5.6) and (2.5.8): a formulation that \cite{ZL08} later adopt. We stress this matter to emphasise that the essential ingredient, on which our solutions build, is itself asymptotic.

We remark briefly on our choice $\ell\gg 1$. As we have already commented, the picture becomes confused after the triggered waves reach the symmetry axis $r=0$. That consideration limited the time over which we reported numerical results. For $\ell=O(1)$, particularly $\ell=1$, because the waves reach $r=0$ on the $O(1)$ rotation time, the mixing of the waves from reflection happens fast. Any ensuing detailed structure suffers considerable internal friction, quickly decays, and is thus hardly visible in the DNS. The interesting features that we find largely pertain to $\ell\gg 1$.

Summarising, our main thrust has been to gain insight about the structures exhibited by the DNS in a simple geometry via the application of asymptotic methods to solve an initial value (itself asymptotic) problem via the LT-method. Our results for the limiting case $E=0$ of \S\ref{no-dis} are robust. As spin-down is a viscous phenomenon, a complete discussion of it requires consideration of finite $E$ solutions. So the comparison in \S\ref{numerics} of the DNS (or rather the FNS) results at $E=10^{-3}$ necessitates use of our approximate theory developed in  \S\ref{dis} for $0<E\ll 1$.

%%%%%%%%%%%%%%%%%%%%%%%%%%%%%%%%%%%%%
%%%%%%%%%%%%%%%%%%%%%%%%%%%%%%%%%%%%%
%%%%%            APPENDICES
%%%%%%%%%%%%%%%%%%%%%%%%%%%%%%%%%%%%%
%%%%%%%%%%%%%%%%%%%%%%%%%%%%%%%%%%%%%

\appendix

\section{An approximate evaluation of the integral (\ref{cut-alternative})\label{Appendix}}

The complication in evaluating (\ref{cut-alternative}), stems from the fact that $\pG$, $\sG$ and $\rho$ (equivalently $q$) defined by (\ref{cut-original}) are complicated functions of $s$; essentially the solution $\pG=\pG^{\pm}(p^{\pm})$ of the cubic (\ref{cut}$d$) is needed (as well as (\ref{cut}$a$)). Nevertheless, whenever $\rho\ll 1$, we may safely neglect $\rho$ in (\ref{cut-original}$c$) to obtain
\be
\label{SS-approx}
\sG^2\,\approx\,s^2\,-\,E(m\pi)^2\,.
\ee
Whence $\rho$, $\pG$ defined by (\ref{cut-original}$a$,$b$), are determined like $\sG$ as functions of $s$ alone. This approximation is equivalent to (\ref{pG-expansion}), correct to $O(E(m\pi)^2)$, under the further approximation $s\ll 1$, valid for $t\gg 1$.

From a more general point of view, when $\rho=O(1)$ the dissipation term $(1-\rho^2)E(m\pi)^2$ in (\ref{cut-original}$c$) is small unless $m$ is large $O(E^{-1/2})$. As noted at the end of \S\ref{cut-section}, when $m=O(E^{-1/2})$  the corresponding $z$-Fourier $m$-mode exists on the Ekman length-scale and is of no interest to us. So in the relevant range $m\ll E^{-1/2}$, the approximation $\sG^2=s^2$ of (\ref{cut-original}$c$) suffices in the construction of $\rho$ and $\pG$ from (\ref{cut-original}$a$,$b$) needed to evaluate (\ref{cut-alternative}).

When $m\rho\gg 1$, which includes the so far undiscussed case $\rho\gg 1$, the Bessel function ratio $\IR_1\bigl(m\pi \rho r\bigr)/\IR_1\big( m\pi \rho\ell\bigr) \approx \exp\bigl(-m\pi\rho(\ell-r)\bigr)$ in the integrand of (\ref{cut-alternative}) deserves further consideration. For $\ell-r\gg (m\pi \rho)^{-1}$ the ratio is negligible; there the integral (\ref{cut-alternative}) essentially vanishes and the formula (\ref{cut-original}$c$) for $\sG$ is irrelevant. Only close to $r=\ell$, where $\ell-r=O\bigl((m\pi \rho)^{-1}\bigr)$, is the ratio finite and hence the integral is finite as well. When $\rho=O(1)$, the neglect of $\rho^2E(m\pi)^2$ relative to  $E(m\pi)^2$ is clearly not justified. This may not matter at lowest order as mentioned in the previous paragraph. Nevertheless, we note that, when $\rho=O(1)$, the distance  $r-\ell$ (now $O((m\pi)^{-1}))\,$) becomes comparable to the $z$-length scale $(m\pi)^{-1}$ of the pertinent $z$-Fourier $m$-mode. Since such length scale comparability is visible in the fan-like structures radiating from the corner $(r,z)=(\ell,0)$, in the $E=0$ results of figures~\ref{fig1-temp} and~\ref{fig2-temp}, the approximation ``might pertain'' to their dissipation visible in the $E=10^{-3}$ results portrayed in figures~\ref{fig1}--\ref{fig4}, panels ($b$), ($e$), ($h$). We write ``might pertain'' as those figures were obtained by a different method, which was explained in \S\ref{no-dis}.

On the integration path of (\ref{cut-alternative}) $s$ is real and so our approximation (\ref{SS-approx}) ensures that, on it, $\sG$ is real too. An explicit form for $\rho$ determined from  (\ref{cut-original}$a$,$b$) is 
\bme
\label{SS-approx-more}
\be
\rho^2\,=\,\dfrac{\iR\sG^2\Sigma^2}{(1+\iR\sG^2/2)^2} \hskip 10mm \mbox{with} \hskip 10mm   \Sigma^2\,=\,1+\dfrac{\iR\sG^2}{4}\,=\,\bigl(\Sigma_+\,+\,\iR\Sigma_-\bigr)^2,
\ee
in which
\be\se
\Sigma
_\pm\,=\,\sqrt{\pm 1\,+\,\sqrt{1+(\sG/2)^4}\,}\Big/{\sqrt2}\,,
\ee
such that
\be\se
\rho^2\,\approx\,\left\{\begin{array}{lll}
\iR \sG^2 \quad& \mbox{for} & \sG\downarrow  0\,,\\[0.2em]
1\quad&  \mbox{as} &  \sG\to \infty\,.\end{array}\right.
\ee
\eme
The fact that the sign of $\rho=\pm\iR^{1/2}\sG(\Sigma_+\,+\,\iR\Sigma_-)\big/(1+\iR\sG^2/2)$ is not unique is of no consequence, as only the dependent term, $\IR_1\bigl(m\pi \rho r\bigr)/\IR_1\big( m\pi \rho\ell\bigr)$, is independent of the sign of $\rho$.

\end{document}